\documentclass[a4paper,fleqn]{cas-sc}

\usepackage[numbers]{natbib}

\def\tsc#1{\csdef{#1}{\textsc{\lowercase{#1}}\xspace}}
\tsc{WGM}
\tsc{QE}
\tsc{EP}
\tsc{PMS}
\tsc{BEC}
\tsc{DE}

\newcommand{\ui}[1]{\textsuperscript{#1}}
\newcommand{\li}[1]{\textsubscript{#1}}

\begin{document}
\let\WriteBookmarks\relax
\def\floatpagepagefraction{1}
\def\textpagefraction{.001}

% Short title
\shorttitle{Effect of annealing on nanoparticle-based CuO--WO$_3$ films}

% Short author
\shortauthors{K.Shaji et~al.}

% Main title of the paper
\title[mode=title]{\texorpdfstring{Thermally-induced microstructural evolution in nanoparticle-based CuO, WO$_3$ and CuO--WO$_3$ thin films for hydrogen gas sensing}{Thermally-induced microstructural evolution in nanoparticle-based CuO, WO3 and CuO--WO3 thin films for hydrogen gas sensing}}                       
% Title footnote mark
% eg: \tnotemark[1]
% \tnotemark[1,2]

% Title footnote 1.
% eg: \tnotetext[1]{Title footnote text}
% \tnotetext[<tnote number>]{<tnote text>} 
% \tnotetext[1]{This document is the results of the research    project funded by the National Science Foundation.}

% \tnotetext[2]{The second title footnote which is a longer text matter to fill through the whole text width and overflow into    another line in the footnotes area of the first page.}

\author[1]{Kalyani Shaji}[type=editor,
    role=Researcher,
    orcid=0009-0008-9320-477X]
% Corresponding author indication
% \cormark[1]

% Footnote of the first author
% \fnmark[1]

\credit{Investigation, Visualization, Writing - Original draft, Writing - Review \& Editing}

% Footnote of the first author
%\fnmark[1]

% Email id of the first author
%\ead{cvr_1@tug.org.in}

% URL of the first author
%\ead[url]{www.cvr.cc, cvr@sayahna.org}

% Second author
\author[1]{Stanislav Haviar}[
    role=Researcher,
    orcid=0000-0001-6926-8927
    ]
    \credit{Investigation, Methodology, Validation, Software, Writing - Original draft}

% Third author
\author[1]{Petr Zeman}[%
   role=Researcher,
   orcid=0000-0001-8742-4487
   ]
\credit{Conceptualization, Validation, Writing - Original draft, Writing - Review \& Editing}

\author[2]{Michal Procházka}[%
   role=Researcher,
   orcid=0000-0003-3364-7006
   ]
\credit{Investigation, Writing - Original draft}

% forth author
\author[1]{Radomír Čerstvý}[%
   role=Researcher,
   orcid=0000-0001-8507-6642
   ]
\credit{Investigation}

% forth author
\author[1]{Nirmal Kumar}[%
   role=Researcher,
   orcid=0000-0003-3739-2565
   ]
\credit{Investigation}

% sixth author
\author[1]{Jiří Čapek}[%
   role=Researcher,
   orcid=0000-0002-3267-7708
   ]
\ead{jcapek@kfy.zcu.cz}
%\ead[URL]{www.sayahna.org}
% Corresponding author indication
\cormark[1]
\credit{Conceptualization, Methodology, Validation, Writing - Original draft, Writing - Review \& Editing, Supervision}

% Address/affiliation
\affiliation[1]{organization={Department of Physics and NTIS - European Centre of Excellence, University of West Bohemia in Pilsen},
    addressline={Univerzitní 8}, 
    city={Plzeň},
    % citysep={}, % Uncomment if no comma needed between city and postcode
    postcode={301 00}, 
    % state={},
    country={Czech Republic}}

% Address/affiliation
\affiliation[2]{organization={New Technologies–Research Centre, University of West Bohemia in Pilsen},
    addressline={Univerzitní 8}, 
    city={Plzeň},
    % citysep={}, % Uncomment if no comma needed between city and postcode
    postcode={301 00}, 
    % state={},
    country={Czech Republic}}

% Corresponding author text
\cortext[cor1]{Corresponding author}
% \cortext[cor2]{Principal corresponding author}

% Footnote text
% \fntext[fn1]{This is the first author footnote. but is common to third author as well.}
% \fntext[fn2]{Another author footnote, this is a very long footnote and it should be a really long footnote. But this footnote is not yet sufficiently long enough to make two lines of footnote text.}

% For a title note without a number/mark
% \nonumnote{This note has no numbers. In this work we demonstrate $a_b$ the formation Y\_1 of a new type of polariton on the interface   between a cuprous oxide slab and a polystyrene micro-sphere placed on the slab.  }

% Here goes the abstract
\begin{abstract}
This study systematically investigates the microstructural evolution of nanoparticle-based CuO, WO$_3$, and composite 'CuO--WO$_3$' thin films induced by their post-deposition annealing. The films were reactively deposited using a magnetron-based gas aggregation technique, with the composite films consisting of alternating monolayers of CuO and WO$_3$ nanoparticles. After deposition, the films were annealed in synthetic air at temperatures ranging from 200 to 400\,$^\circ$C and characterized using scanning electron microscopy, X-ray diffraction, Raman spectroscopy, and X-ray photoelectron spectroscopy. Annealing of the CuO films led to the most pronounced changes associated with a gradual enhancement of crystallinity accompanied by significant particle growth with increasing annealing temperature, while the WO$_3$ and CuO-WO$_3$ films were more thermally stable to crystallization and particle growth. Notably, at 400\,$^\circ$C, the CuO--WO$_3$ films crystallized into a novel $\gamma$-CuWO$_4$ phase. The annealed films were further evaluated for their gas-sensing performance upon H$_2$ exposure and the obtained results were analyzed in relation to film properties and the microstructural evolution induced by annealing.

\end{abstract}

% Use if graphical abstract is present
% \begin{graphicalabstract}
% \includegraphics{figs/grabs.pdf}
% \end{graphicalabstract}

% Research highlights
\begin{highlights}

\item Reactive synthesis of nanoparticle-based CuO,WO$_3$ and CuO-WO$_3$ composite films using a magnetron-based gas aggregation source.

\item Systematic investigation of the films' microstructural evolution during annealing in synthetic air up to 400\,$^\circ$C.

\item Novel metastable $\gamma$-CuWO$_4$ phase identified upon annealing the composite film at 400\,$^\circ$C.

\item Gas-sensing response of the films to H$_2$ exposure analyzed and discussed in relation to their thermally-induced changes.

\end{highlights}

% Keywords
% Each keyword is seperated by \sep
\begin{keywords}
CuO, WO$_3$, CuO--WO$_3$ \sep Microstructural evolution \sep Thermal annealing \sep  Nanoparticle-based thin films \sep Gas aggregation source \sep Conductometric gas sensors \sep Hydrogen gas sensing 
\end{keywords}

\maketitle

\section{Introduction}
\label{intro}

Hydrogen has gained significant interest as a clean and renewable energy carrier, which makes it a promising candidate for various applications, including power generation, large-scale energy storage, and chemical manufacturing \cite{Manoharan2019}, \cite{Muradov2008}. However, its widespread applicability is hindered by safety concerns due to its low flammability limit (4\% in air) and extremely low ignition energy (<0.02\,mJ), which present substantial explosion risks \cite{Okazaki2003}. Furthermore, delicate devices such as fuel cells require precise control of low hydrogen flow rates to ensure optimal performance. As a result, the development of reliable hydrogen sensing technologies is essential for monitoring and regulating hydrogen concentrations across various applications, ensuring both safe and efficient operations.

Among gas-sensing technologies, conductometric sensors have attracted considerable interest for their high sensitivity, fast response, and cost-effectiveness \cite{Antonio2022}, \cite{Joshi2018}. These sensors operate on the principle that the electrical conductivity of the sensing material changes in response to the target gases (analytes), which enables their detection and quantification. In particular, metal oxide semiconductors (MOS) are renowned for their superior performance as conductometric sensors \cite{Goel2023, Nadargi2023}. In this case, gas species from the ambient atmosphere adsorb and desorb on the semiconductor surface, which alters the carrier concentration near the surface (known as the depletion layer (DL) for n-type MOS) and, in turn, the electrical properties of the material. Pure MOS sensors (not enhanced by noble-metal catalysts) are typically operated at elevated temperatures (200 -- 400\,$^{\circ}$C) to facilitate the dissociation of H$_2$ molecules on the surface, which is a critical process for the sensing mechanism.

One of the key factors in the gas sensing performance is the microstructure of the sensing material, as it directly influences interactions between the sensing material and the analyte \cite{Zhao2023}. From this point of view, nanostructural materials are of particular interest. First, their high surface area significantly increases the number of active sites for gas adsorption and subsequent reactions. Second, optimizing the dimensions and spatial arrangement of the individual building blocks (e.g., nanoparticles (NPs), nanotubes) within the nanostructural materials significantly enhances the ability (known as transducer function) to convert chemical interactions with the analyte gases into a measurable electrical signal \cite{Girolamo2009, Ghenadii}. Third, so-called ‘necks’ can be formed at the junctions between the individual building blocks of the material, for example, when the material is exposed to an elevated temperature. These 'necks' significantly affect the changes in the measured conductivity of the percolation network in the material depending on whether the material is exposed to the analyte \cite{Jun2009}. The reason is that smaller ‘necks’ can be ‘closed’ or ‘open’ for the electrical current, depending on the dimensions of the DL formed. Fourth, combining building blocks of different metal oxides may lead to the formation of heterojunctions at their interfaces. These heterojunctions are typically highly sensitive to the analyte and can lead to the formation of even deeper DL. This, in turn, amplifies changes in the electrical conductivity of the percolation network, since even larger ‘necks’ can be ‘closed’ or ‘open’ during the sensing process \cite{Korotcenkov2017, Yang2021}. 

Among MOS, those composed solely of CuO or WO$_3$ NPs have shown excellent potential for gas-sensing applications \cite{Steinhauer2021, Dong2020}. In our recent works \cite{ Kumar2020, Kumar2021, Haviar2018}, we also demonstrated that the sensing performance can be further enhanced by the synergistic combination of CuO NPs with a WO$_3$ thin film due to the formation of nano-sized p-n heterojunctions between p-type CuO and n-type WO$_3$. Building on these results, we have optimized our deposition technique, which uses a magnetron-based gas-aggregation source (GAS), for the one-step controlled synthesis of the composite thin films consisting of alternating monolayers of CuO and WO$_3$ NPs \cite{Shaji2024}. This physical deposition technique, leveraging low-temperature discharge plasma, allows us to prepare high-purity NPs with tunable size, composition, and morphology, and further explore the effect of a composite of CuO and WO$_3$ NPs on hydrogen gas sensing.

Given that the microstructure of nanoparticle-based (NP-based) thin films is of key importance in determining their gas-sensing performance, we conducted a comprehensive multi-scale structural study (from atomic to morphological scale), which is complemented with hydrogen-sensing measurements. In particular, we systematically investigated the effects of post-deposition thermal annealing in air at temperatures between 200 -- 400\,$^{\circ}$C (mimicking also operation of the MOS sensors at elevated temperatures) on the microstructure evolution of NP-based CuO, WO$_3$, and CuO--WO$_3$ thin films with the aim to further optimize the properties of these films and achieve a deeper understanding of the thermally-induced mechanisms that govern their gas-sensing behavior. 

\section{Experimental details}
\subsection{Film deposition and post-deposition annealing}

The thin films investigated in this study were prepared using a custom-built deposition system consisting of a GAS source mounted on a DN200ISO-K six-way cross vacuum chamber equipped with a rotating substrate holder and load-lock system. The GAS (Nanogen-Trio, Mantis deposition) consisted of three 1'' magnetrons embedded into an axially movable holder of magnetrons installed within a grounded cylindrical aggregation chamber terminated by a domed ending with an exit orifice of 4\,mm in diameter. In this work, just two magnetrons equipped with high-purity (at least 99.95\%) 3.2\,mm thick Cu and W targets were used. Each magnetron was driven by a DC power supply operated at a constant discharge power of 25\,W. A quartz crystal microbalance (QCM) mounted on a motorized shutter was used to measure the mass ﬂux of NPs before and during the deposition process. Our in-house-built LabVIEW-based software effectively orchestrated the power supplies, ﬂow controllers, and motorized QCM, ensuring reliable process repeatability. More details about the deposition process can be found in Ref.~\cite{Shaji2024}.  For this work, individual NP-based thin films composed exclusively of CuO and WO$_3$, as well as complex composite thin films with alternating monolayers of CuO and WO$_3$ NPs, were prepared. Note that from here on, we denote the individual NP-based thin films of the respective material as CuO or WO$_3$, and the composite films of alternating CuO and WO$_3$ monolayers as CuO--WO$_3$. 

To explore the effects of annealing of the films, six samples each of CuO, WO$_3$, and CuO--WO$_3$ composite were prepared. The films were deposited onto polished and ultrasonically-cleaned $5\times15$\,mm$^2$ Si (100) substrates. To ensure uniformity of the film across the entire substrate area, the depositions were divided into two sets, with three substrates per deposition. One sample each from CuO, WO$_3$, and CuO--WO$_3$ was kept aside as the as-deposited, while the other five were separately annealed to temperatures 200, 250, 300, 350, and 400\,$^{\circ}$C which corresponds to the typical operating temperatures of most MOS-based conductometric sensors. The films were annealed in a furnace (Clasic 1800) under synthetic air conditions at atmospheric pressure for 6 hours, followed by comprehensive characterization. First XRD, Raman spectroscopy, and SEM top-view analyses were carried out. Subsequently, each sample was pre-scratched and carefully broken into two pieces, each measuring 5$\times$7.5 mm$^2$, with one part subjected to XPS study and the other to SEM cross-section analysis. 

\subsection{Film characterization}
\label{charac}

The scanning electron microscope (SEM) (Hitachi SU 70) was used to make top-view and cross-sectional images of the as-deposited and annealed samples and investigate the effect of heating on their microstructure. A primary energy of 10\,keV and 15\,keV was used for cross-sectional and top-view imaging, respectively. Top-view SEM images were analyzed using the ImageJ software to determine NP sizes. Approximately 50 particles, where the boundaries were reasonably visible, were selected and their diameters were measured manually with the imaging tools. The size distribution was then plotted as a histogram and fitted with log-normal distribution to estimate the average particle diameter. From the cross-sectional micrographs, the thickness of the as-deposited films was estimated to be $300\pm10$ nm.

The crystallographic structure of the prepared films was examined by glancing incidence X-ray diffraction utilizing a diffractometer (X’Pert PRO MPD, PANalytical), operating at an accelerating voltage of 40\,kV and a tube current of 40\,mA, using CuK$\alpha$ ($\lambda$ = 0.154187\,nm) at a glancing angle of 0.6\,$^{\circ}$. Diffraction patterns were obtained by continuous scanning in the 2$\theta$ range of 10-80\,$^{\circ}$ with a step size of 0.05\,$^{\circ}$ and time per step of 12\,s. Data analysis was conducted using the PANalytical software package, HighScore Plus. The phase composition was also studied with Raman spectroscopy (LabRAM HR Evolution, Horiba Jobin Yvon ) using a 532\,nm laser.

X-ray photoelectron spectroscopy (XPS) measurements were performed in an ultrahigh vacuum chamber with a base pressure of $\leq$3 x 10$^{-8}$\,Pa, using a non-monochromatic X-ray source XR 50 operated with Mg K$\alpha$ line (h$\nu$ = 1253.6\,eV) and the hemispherical analyzer (Phoibos 150, SPECS Surface Nano Analysis GmbH) with a multichannel CMOS detector. The survey spectra were obtained by 5 scans with an energy step size of 0.5\,eV and pass energy of 50\,eV. Core-level spectra were measured by 20 scans with an energy step size of 0.05\,eV and pass energy of 30\,eV. Samples were mounted on Ti or Mo sample holders using conducting silver paste. For analyzing the measured spectra, the KolXPD software was used. All spectra have been charge corrected according to adventitious C~1s spectral component (C–C, C–H) with a binding energy (BE) of 284.8\,eV \cite{Biesinger2022}. The peaks were fitted using Shirley background and Voigt and Voigt doublet functions for deconvolution.

\subsection{Measurement of gas sensing properties}

Gas sensing measurements were performed on 200\,nm thick films deposited onto 9\,$\times$\,9\,mm$^2$ quartz glass substrates with square 2$\times$2 mm$^2$ electrodes in the corners comprising sputter-deposited 50\,nm Cr layer followed by 100\,nm Pt layer.

A custom-built sensor testing station was developed to assess the H$_2$ gas sensing performance of the annealed samples using the four-point probe Van der Pauw method. Stable electrical contacts were established using platinum-coated contact pointy clamps arranged in a square pattern with a spacing of 8\,mm, which were pressed onto the sample electrodes. This whole setup was inside a brass chamber with a total volume of 3\,cm$^3$. The sample was heated using a ceramic hot plate (Bach Resistor Ceramics GmbH, Germany), and a thin thermocouple (0.3\,mm in diameter) was attached to the top surface of the sample to monitor its surface temperature. More details can be found in Ref.  \cite{Haviar2025}.

The testing atmosphere was precisely regulated using three mass flow controllers (Alicat Scientific Ltd.) for the N$_2$, O$_2$, and H$_2$ gases. To evaluate the H$_2$ response, the sample was exposed to alternating 5-minute pulses of 1\% H$_2$ by vol. in synthetic air while maintaining the total flow rate at 100\,sccm. Resistance measurements were performed using a precise current source (Keithley 6220), and two voltmeters (Keithley 6514). 
Each sample was subjected to a standardized testing protocol. Prior to the measurements, the sample was pre-heated to 150\,$^{\circ}$C in a constant flow of synthetic air and maintained at this temperature for one hour to stabilize the resistance values. Subsequently, the gas sensing response was recorded for selected annealed samples at operating temperatures ranging from 200 to 400\,$^{\circ}$C.

\section{Results}
\label{Results}

\subsection{SEM}
\label{SEM}

\begin{figure}[h]
    \centerline{\includegraphics[width=120mm]{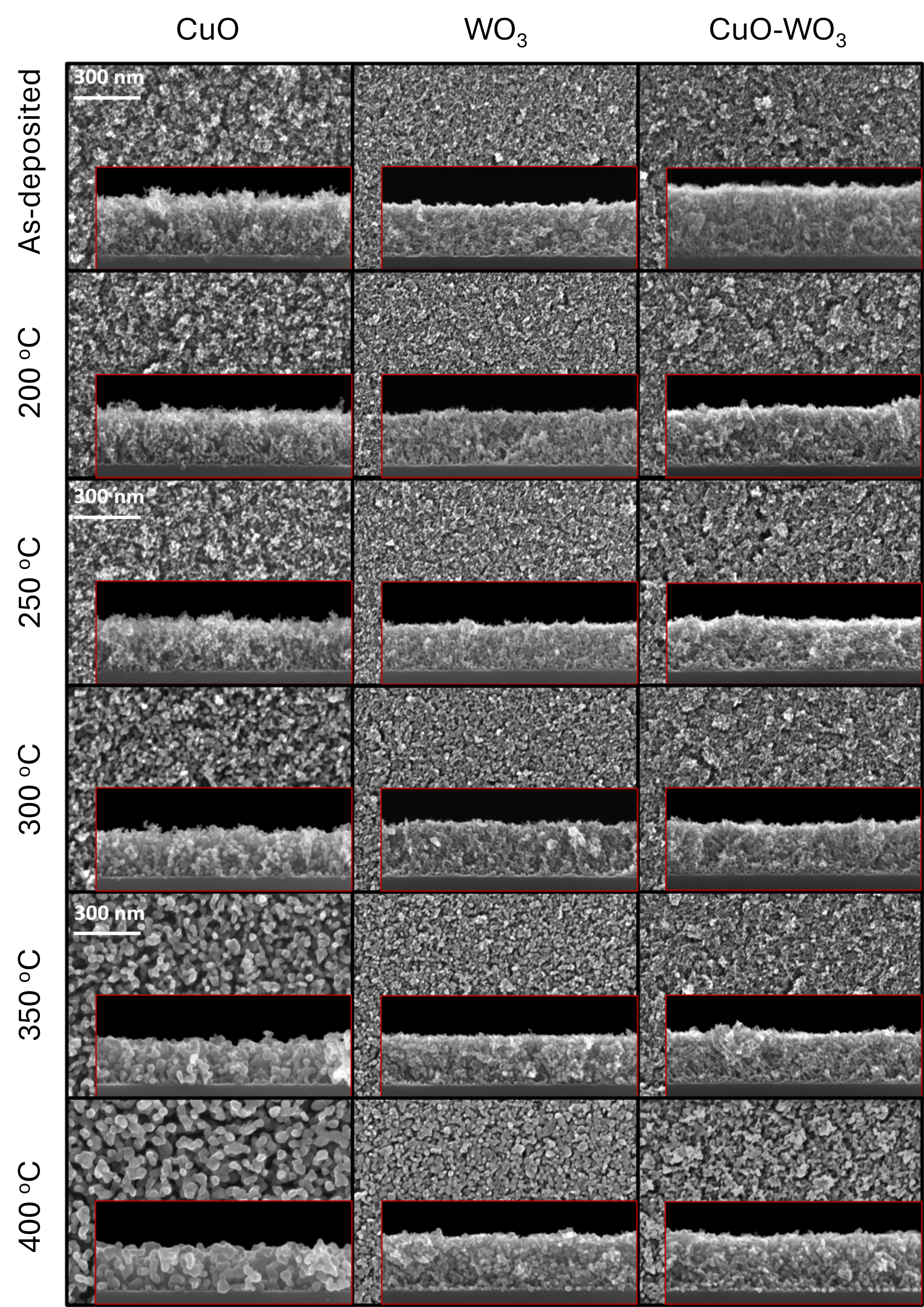}}
            \caption{SEM micrographs of top-views of the NP-based CuO, WO$_3$, and CuO--WO$_3$ films for the as-deposited state and after annealing at different temperatures. The insets show the cross-sectional micrographs of the respective samples with the same magnification as for the top-views.}
            \label{fig SEM}
\end{figure}

The top-view SEM micrographs, which illustrate the structural evolution of NP-based thin films of CuO, WO$_3$ and their CuO--WO$_3$ composite as a function of the annealing temperature, are shown in Fig.\ref{fig SEM}. The corresponding cross-sectional views are presented as insets.

The films in their as-deposited state have already been thoroughly examined in our previous work \cite{Shaji2024}. The films consist of spherical NPs with a mean diameter of approximately 9\,nm for both materials with corresponding Full Width at Half Maximum (FWHM) values of 1.5 and 3\,nm for CuO and WO$_3$ NPs, respectively. These values were determined by analyzing top-view SEM micrographs of isolated NPs dispersed on the substrate. However, this approach could not be applied to the annealed films in the present study, as the focus here is on the interactions between touching NPs during the annealing. For this reason, we utilized the ImageJ software package to estimate the dimensions of NPs from the SEM micrographs of the annealed films, as described in detail in Section \ref{charac}. Although individual NPs with clearly distinguishable boundaries were selected manually from the SEM micrographs, the limitation in the resolution of objects at the nanoscale does not prevent the possibility of selecting agglomerates (loosely bonded NPs) and/or aggregates (fused or strongly bonded NPs forming a new particle) instead of individual NPs. Consequently, the measured size may be larger than their real dimensions. Nevertheless, calculating the mean diameter still provides a reasonable approximation of the overall particle size.

Considering the microstructure of the as-deposited films, the WO$_3$ film exhibits a less porous microstructure compared to the CuO film, despite the fact that NPs are of the same size. As expected, the porosity of the composite film appears to be between that of the WO$_3$ film and the CuO film, but the microstructure is more similar to that of the WO$_3$ film.

The microstructure of the CuO films remains unchanged up to 250\,$^\circ$C when annealed. A noticeable increase in particle size begins at 300\,$^\circ$C, with significantly larger NPs observed at 350 and 400\,$^\circ$C. Based on our analysis of the SEM micrographs, the mean diameter of the NPs was derived from the fitted log-normal distributions, yielding values of 11, 19, and 27\,nm for the films annealed at 300, 350, and 400\,$^\circ$C, respectively, with corresponding FWHM values of  8, 8, and 15\,nm. 

For the WO$_3$ films, the microstructure remains stable up to 300\,$^\circ$C. A slight increase in the diameter of the NPs is observed at 350\,$^\circ$C. This increase becomes more pronounced at 400\,$^\circ$C, where the mean diameter of NPs is 12\,nm (FWHM=7\,nm). These findings align with previous studies, which report a more pronounced growth of WO$_3$ NPs at temperatures exceeding 600\,$^\circ$C \cite{Chai2015}. 

Finally, the microstructure of the CuO--WO$_3$ composite film remains stable up to 350\,$^\circ$C, which suggests that the presence of both constituents in the film contributes to its structural stability. This stabilization is notable, as the microstructures of the individual constituents are stable only to lower temperatures (250\,$^\circ$C for the CuO NPs and 300\,$^\circ$C for the WO$_3$ NPs). Annealing at 400\,$^\circ$C results in a slight increase in particle size with a mean diameter of approximately 11\,nm (FWHM=6\,nm).

\subsection{XRD}
\label{XRD}

\begin{figure}[h]
    \centerline{\includegraphics[width=1\linewidth]{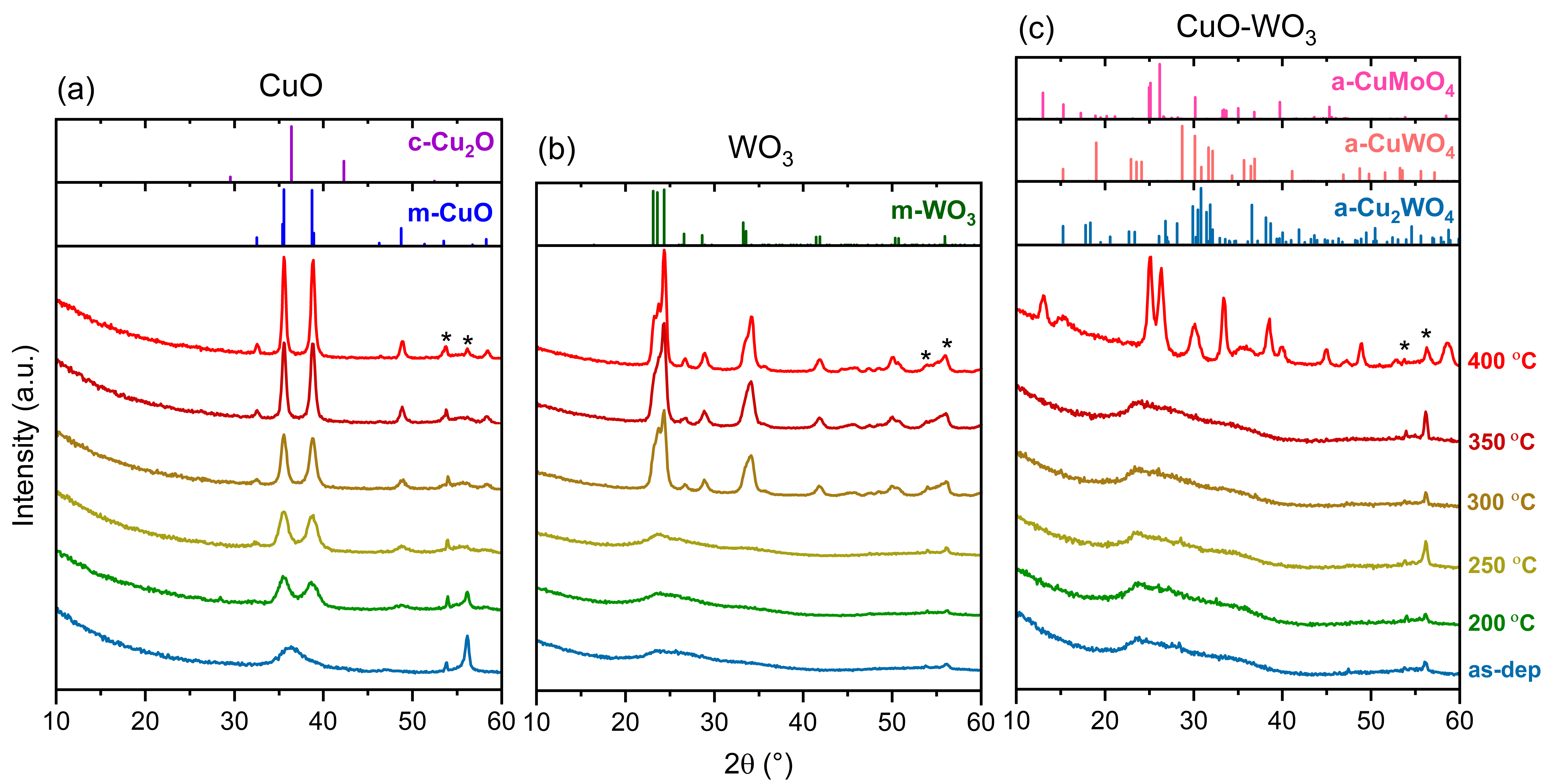}}
            \caption{XRD patterns of NP-based CuO (a), WO$_3$ (b), and CuO--WO$_3$ (c) films for the as-deposited state and after annealing at different temperatures. The diffraction peaks labeled with an asterisk '*' in the XRD patterns can be attributed to the silicon substrate.}
            \label{fig:XRD 400C}
\end{figure}

Fig. \ref{fig:XRD 400C} shows the XRD patterns of the as-deposited films and the films annealed up to 400\,$^{\circ}$C.
From Fig. \ref{fig:XRD 400C}a, which corresponds to the XRD pattern of CuO films, it is evident that the annealing process has an influence on the intensity and the width of diffraction peaks. The as-deposited film shows a broad peak indicating a low crystallinity of the film. The peak position overlaps with the positions of the powder diffraction standards of monoclinic CuO (m-CuO; PDF Card No. 00-048-1548) and cubic Cu$_2$O  (c-Cu$_2$O; PDF Card No. 05-0667), indicating the presence of both phases in the as-deposited film. When the film is annealed to 200\,$^{\circ}$C, two distinct diffraction peaks emerge at 2$\theta$ values of about 35.5$^{\circ}$ and 38.7$^{\circ}$. These peaks can be assigned exclusively to m-CuO. The first peak corresponds to (002) and (11$\Bar{1}$) diffraction doublet, while the second corresponds to (111) and (200) doublet. With increasing annealing temperature, these peaks become narrower and more intense, which suggests the growth of crystallites and improvement in the crystallinity of the film, respectively. The size of coherently diffracting domains was calculated from the FWHM of the (111) peak using the Scherrer equation, yielding values of 5, 7, 12, 18, and 23\,nm for the films annealed at temperatures of 200, 250, 300, 350 and 400\,$^{\circ}$C, respectively. Let us note that these values are in a good agreement with the size of the CuO NPs determined from the SEM micrographs (see Sec. \ref{SEM}). 

The XRD patterns of the WO$_3$ films in Fig. \ref{fig:XRD 400C}b exhibit very broad diffraction peaks from the as-deposited state up to an annealing temperature of 250\,$^{\circ}$C. Such peaks are characteristic of a disordered structure. At these relatively low temperatures, the thermal energy is insufficient to promote a significant crystallization or crystallite growth \cite{Chai2015, CHEN2021}. At 300\,$^{\circ}$C, the XRD pattern contains distinct diffraction peaks at 2$\theta$ values of 23.25, 23.75, and 24.35$^{\circ}$. These peaks correspond to the (002), (020), and (200) lattice planes of monoclinic WO$_3$ (m-WO$_3$; PDF Card No. 01-083-0950), respectively. At higher temperatures, the intensity of the (200) peak increases, which is attributed to improving crystallinity. However, the changes in FWHM are too small to draw definitive conclusions about the crystallite growth.

The as-deposited CuO--WO$_3$ composite film, consisting of alternating CuO and WO$_3$ monolayers, exhibits an XRD pattern (Fig. \ref{fig:XRD 400C}c) characterized by broad peaks similar to those of the individual CuO and WO$_3$ films. Its disordered state is preserved up to 350\,$^{\circ}$C, which is a temperature higher than the crystallization thresholds of both binary films. Notably, further annealing to 400\,$^{\circ}$C leads to the emergence of a novel phase, which we refer to as $\gamma$-CuWO$_4$. Its XRD pattern does not show conclusively peaks corresponding to CuO, WO$_3$, or any of their compounds recorded in the PDF database (e.g., triclinic CuWO$_4$ (a-CuWO$_4$); PDF Card No. 04-009-6293 or triclinic Cu$_2$WO$_4$ (a-Cu$_2$WO$_4$); PDF Card No. 00-041-0948). Instead, it closely resembles to the known triclinic $\gamma$-CuMoO$_4$ phase (PDF Card No. 04-009-2227). In our parallel study, we used ab initio simulations \cite{houska2025newpolymorphgammaCuWO4inspired} and we confirm the existence of this phase as a metastable form that might be stabilized by its lower density and/or Cu-rich composition.

\subsection{Raman spectroscopy}
\label{Raman}

\begin{figure}[h]
    \centerline{\includegraphics[width=1\linewidth]{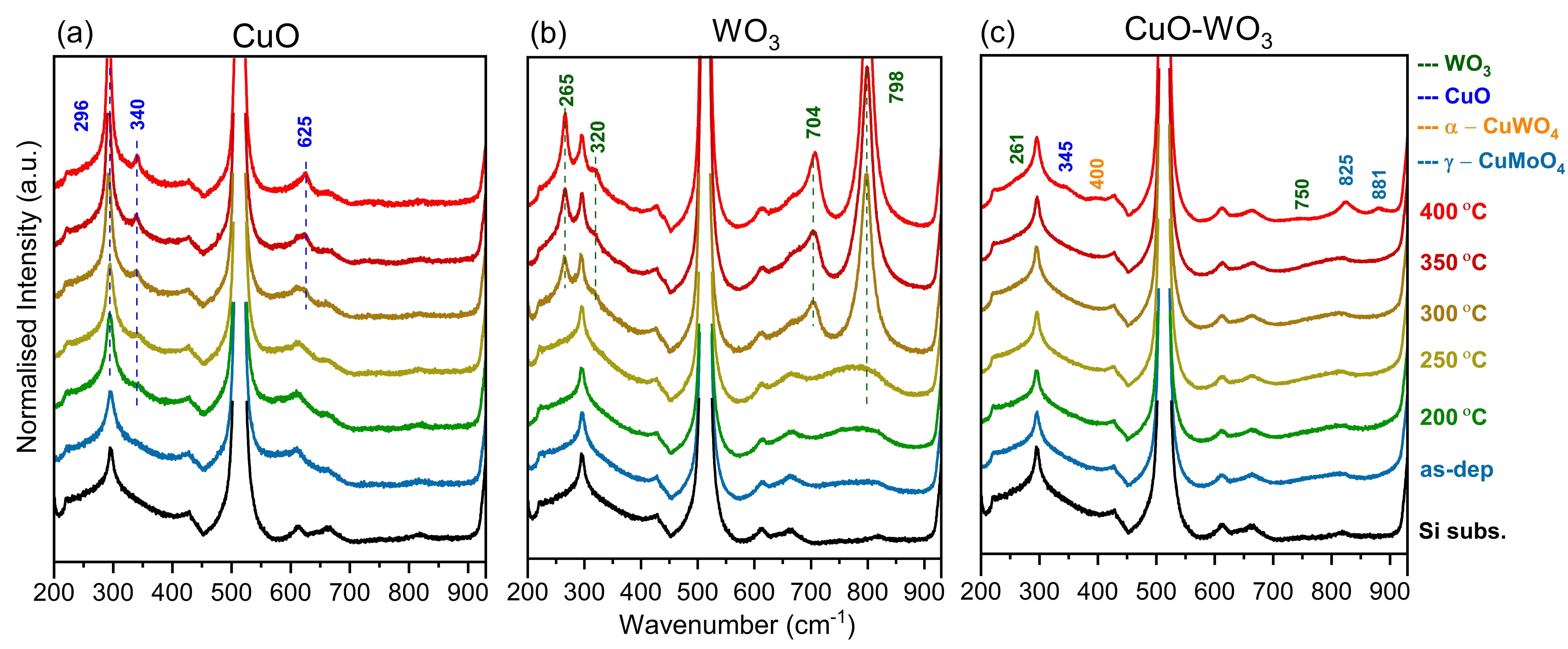}}
            \caption{Raman spectra of NP-based CuO (a),  WO$_3$ (b), and  CuO--WO$_3$ (c) films for the as-deposited state and after annealing at different temperatures. The spectrum corresponding to the bare Si substrate is shown as well as a reference.}
            \label{fig:Raman 400C}
\end{figure}

Fig. \ref{fig:Raman 400C} presents Raman spectra of the as-deposited and annealed films. The Raman spectrum of the Si (100) substrate is shown at the bottom for reference. 

From the Raman spectra of CuO films (Fig. \ref{fig:Raman 400C}a), a peak overlapping with the Si vibrational mode is visible at 296\,cm$^{-1}$, which corresponds to the A$_g$ mode of m-CuO, linked to phase rotations \cite{Akgul2014}. With heat treatment, there is an increase in the relative intensity of this peak to the intensity of the Si peak at around 513\,cm$^{-1}$, which confirms that the changes of the peak at 296\,cm$^{-1}$ can be associated with CuO. A second, weak peak is observed at 340\,cm$^{-1}$, which gradually intensifies with increasing temperature. Additionally, a peak at 625\,cm$^{-1}$ may indicate the presence of another vibrational mode of CuO due to its asymmetrical nature. Both peaks correspond to the B$_g$ modes of m-CuO, with the 340\,cm$^{-1}$ peak attributed to Cu–O bending and the 625\,cm$^{-1}$ peak associated with symmetric oxygen stretching \cite{Akgul2014}. The sharper and more intense Raman peaks observed at higher temperatures can be attributed to the substantial increase in particle size of CuO , as discussed in Section~\ref{SEM} \cite{Akgul2014, Xu1999}.

The Raman spectra of the WO$_3$ films  (Fig. \ref{fig:Raman 400C}b) in both the as-deposited state and after annealing up to 250\,$^{\circ}$C show a broad, asymmetric peak in the range of 700–800\,cm$^{-1}$. This broad feature may indicate a low crystallinity of the film \cite{Nanba1991, Santato2001}, which is consistent with the XRD results (Fig. \ref{fig:XRD 400C}b) that show a disordered state below 300\,$^{\circ}$C. Upon annealing to 300\,$^{\circ}$C, a sharp peak appears at 798\,cm$^{-1}$ along with weak peaks at 265 and 704\,cm$^{-1}$. A very weak peak at about 320\,cm$^{-1}$ also emerges at 300\,$^{\circ}$C. These bands (at 265, 320, 704, and 798\,cm$^{-1}$) closely align with the wavenumbers of the four strongest modes of m-WO$_3$ \cite{Santato2001}. The bands at 265 and 320\,cm$^{-1}$ correspond to the O–W–O bending modes of the bridging oxygen, while the bands at 704 and 798\,cm$^{-1}$ correspond to the stretching modes \cite{Santato2001, Song2019}. With further heat treatment, the peaks become sharper and more intense. Unlike CuO, these changes are not linked to particle growth, as the size of WO$_3$ NPs remains nearly unchanged after annealing (see Fig.\ref{fig SEM}c), but they are attributed to enhanced crystallization.

In the CuO--WO$_3$ composite films, the Raman spectra from the as-deposited state up to 350\,$^{\circ}$C are characterized by broad peaks, while distinct peaks appear only after annealing at 400°C (Fig. \ref{fig:Raman 400C}c). This observation aligns well with the XRD findings (Fig. \ref{fig:XRD 400C}c). Although the peak identification cannot be fully conclusive, the peaks can be attributed to the CuO, WO$_3$ and  CuWO$_4$ phases. The weak signal at 261\,cm$^{-1}$ and the broad band observed around 750\,cm$^{-1}$(likely deconvoluted into two prominent peaks at 704 and 798\,cm$^{-1}$) correspond to m-WO$_3$, while the weak peak at 345\,cm$^{-1}$ can be attributed to m-CuO. Other Raman peaks detected at 400\,cm$^{-1}$, 825\,cm$^{-1}$ and 881\,cm$^{-1}$ appear to be associated with CuWO$_4$. The peak at 400\,cm$^{-1}$ is commonly observed in the stable triclinic CuWO$_4$ phase (refered to here as $\alpha$-CuWO$_4$) \cite{Ruiz‐Fuertes2008}.  The peaks at 825\,cm$^{-1}$ and 881\,cm$^{-1}$ closely match the characteristic positions of $\gamma$-CuMoO$_4$ nanostructures studied by Ali et al. \cite{Ali2022}. Specifically, the Raman spectra of $\gamma$-CuMoO$_4$ show peaks at 811 and 837\,cm$^{-1}$ along with a prominent peak at 880\,cm$^{-1}$. Since no Mo was detected in the composite film, and the novel $\gamma$-CuWO$_4$ phase was revealed by XRD and further confirmed by ab initio simulations as a structural prototype of  $\gamma$-CuMoO$_4$ \cite{houska2025newpolymorphgammaCuWO4inspired}, we suggest that the peak at 825\,cm$^{-1}$ in the composite film may result from a combination of the 811 and 837\,cm$^{-1}$ modes, while the peak at 881\,cm$^{-1}$ is consistent with the peak at 880\,cm$^{-1}$. This alignment indicates that these peaks likely represent vibrational modes equivalent to the $\gamma$-CuMoO$_4$ phase.

\subsection{XPS}
\label{XPS}

\begin{figure}[h]
    \centerline{\includegraphics[width=1\linewidth]{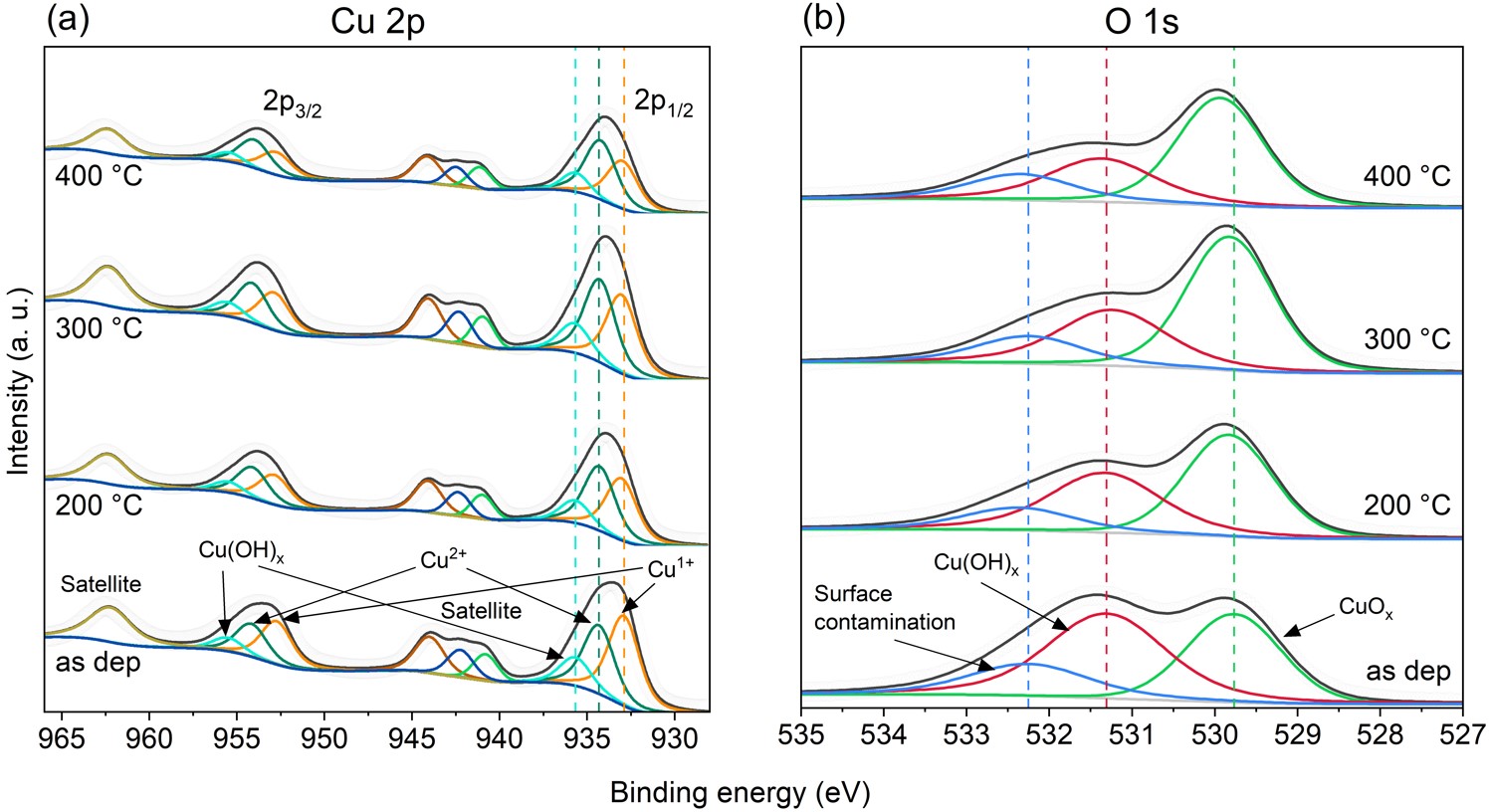}}
            \caption{XPS core-level spectra of Cu 2p (a) and  O 1s (b) of NP-based CuO film for the as-deposited state and after annealing at different temperatures. The dotted lines indicate the respective peak positions in the as-deposited state.}
            \label{fig XPS CuO}
\end{figure}

\begin{figure}[h]
    \centerline{\includegraphics[width=1\linewidth]{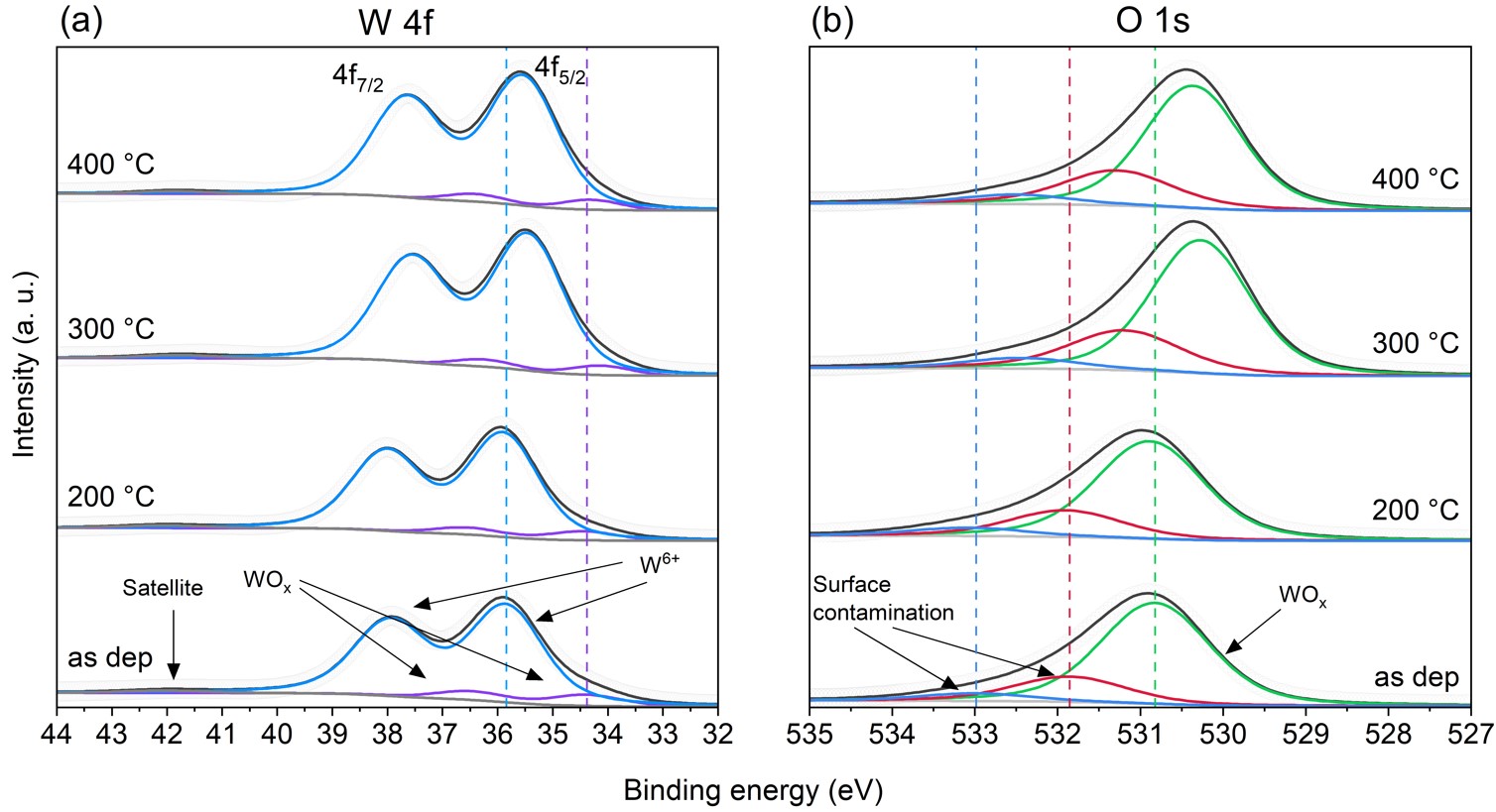}}
            \caption{XPS core-level spectra of W 4f (a) and O 1s (b) of NP-based WO$_3$ film for the as-deposited state and after annealing at different temperatures. The dotted lines indicate the respective peak positions in the as-deposited state.}
            \label{fig XPS WO3}
\end{figure}

\begin{figure}[h]
    \centerline{\includegraphics[width=1\linewidth]{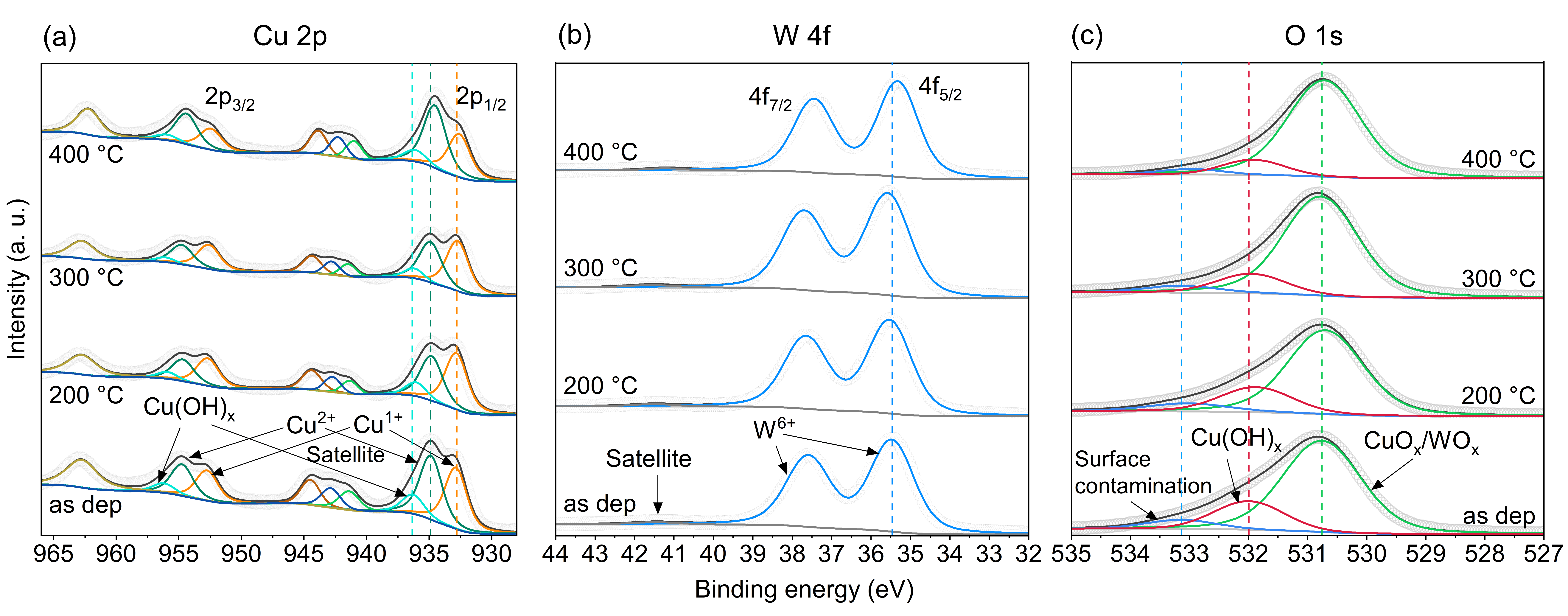}}
            \caption{XPS core-level spectra of Cu 2p (a), W 4f (b), and O 1s (c) of NP-based CuO--WO$_3$ film for the as-deposited state and after annealing at different temperatures. The dotted lines indicate the respective peak positions in the as-deposited state.}
            \label{fig XPS mix}
\end{figure}

XPS analysis was conducted to investigate the evolution of the bonding states in the films with increasing annealing temperature. Note that the surface of the composite film has been topped by a layer of the CuO NPs with 50$\%$ coverage of the underlying WO$_3$ monolayer, instead of the complete coverage by the CuO monolayer (10 nm). This approach ensured that comprehensive information is obtained via XPS which has a characteristic penetration depth below 10\,nm. The analyzed core-level spectra of O, Cu, and W are summarized in  Figs. \ref{fig XPS CuO}, \ref{fig XPS WO3} and \ref{fig XPS mix}.

%CuO NPs as-deposited%
In the Cu\,2p spectrum of the as-deposited CuO film (Fig. \ref{fig XPS CuO}a), it is possible to identify a component at 932.89\,eV, likely corresponding to a mixture of Cu\ui{1+} and metallic Cu\ui{0} states, \cite{Biesinger2011, Moulder1992}. The second peak at 934.33\,eV corresponds to Cu\ui{2+} \cite{Biesinger2011, Moulder1992}. The position of the Cu\ui{2+} is slightly shifted compared to most references, but the split of Cu\ui{2+} and Cu\ui{1+} states is in the reasonable range (here 1.43\,eV).  The remaining features in the spectrum include typical satellite peaks of CuO and Cu(OH)$_x$. The O\,1s spectrum (Fig. \ref{fig XPS CuO}b) consists of three components: the one at 529.77\,eV represents both copper oxides, Cu$_2$O and CuO. The other two components at 531.23\,eV and 532.07\,eV correspond to surface contamination such as Cu(OH)$_x$, other hydroxides, or carbonates.%
%CuO NPs with temperature%
With annealing, we can observe that the intensity of the contamination-related peaks decrease due to desorption while that of the copper oxide peaks increase. This trend is evident in both Cu\,2p and O\,1s spectra.
Additionally, there is an increase in the relative abundance of the Cu\ui{2+} state at the expense of Cu\ui{1+}, which indicates oxygen saturation and the formation of the CuO phase. At the same time, while it is not possible to confirm or deny a metallic component Cu\ui{0}, from the strong diffraction peaks of m-CuO at elevated temperatures (Fig. \ref{fig:XRD 400C}a), one can assume that any metallic state present in the as-deposited film likely vanishes during annealing, if it existed at all. 

%WOx NPs as-deposited%
The XPS spectra of WO$_3$ films are shown in Fig. \ref{fig XPS WO3}. In the W\,4f spectra (Fig. \ref{fig XPS WO3}a), a major component at 35.84\,eV can be identified as W\ui{6+}, which is 
consistent with literature values of 35.7\,eV \cite{Shpak2007} and 35.8\,eV \cite{Moulder1992}. A minor component at 34.38\,eV is attributed to an unspecified substoichiometric oxide state \cite{Moulder1992, Khyzhun2002}. With increasing annealing temperature, a small but continuous decrease in intensity of this peak is observed, indicating further oxidation. Fig. \ref{fig XPS WO3}b shows the evolution of the O\,1s peak in the WO$_3$ films. Similar to the W\,4f spectrum, the main peak at 530.82\,eV corresponds to the tungsten oxides \cite{Moulder1992, Shpak2007}. The other two components at higher binding energies are attributed to surface contamination, likely originating from hydroxides, carbonates, or other surface-bound species.
%WOx NPs with temperature%
With increasing temperature, in addition to oxidation toward fully stoichiometric WO$_3$, the W\ui{6+} peak shifts to lower binding energies above the temperature of 300\,$^{\circ}$C. This shift may be due to a microstructural evolution and crystallization of WO$_3$, as clearly evident from XRD pattern for 300\,$^{\circ}$C in Fig. \ref{fig:XRD 400C}b. A similar phenomenon was reported by Liu et al. \cite{Liu2018} after TiO$_2$ crystallization during annealing at 400\,$^{\circ}$C.

%Mixed particles
Fig. \ref{fig XPS mix} shows XPS results for the CuO--WO$_3$ composite film. Similar to the CuO film, three components can be identified: Cu\ui{1+}(+Cu\ui{0}) at 932.86\,eV, Cu\ui{2+} at 934.86\,eV, and Cu(OH)$_x$ at 936.30\,eV. In contrast to the CuO film, the composite films exhibit a larger peak for the less oxidized copper states compared to the Cu\ui{2+} state. This can be explained by a preferential binding of the available oxygen to W, as can be seen in the W\,4f spectra in Fig. \ref{fig XPS mix}b, where only a single component W\ui{6+} is identified at 35.47\,eV. A sudden change in the Cu\,2p component ratio is visible at 400\,$^{\circ}$C. Here, CuWO\li{4} is formed, as evident in the XRD pattern in Fig. \ref{fig:XRD 400C}c. Studies in Refs. \cite{Gajraj2021, Khyzhun2005} supports the potential formation of CuWO$_4$ with the Cu\ui{2+} state appearing at the same position as in CuO. 
Therefore, the peak at 934.86\,eV likely represents both CuO and CuWO\li{4}. The same applies to the W\,4f spectra, where the only spectra component represents both WO\li{3} and CuWO\li{4}.
Fig. \ref{fig XPS mix}c shows O\,1s spectra of the composite film. The main peak at 530.76\,eV represents the states of all the oxides mentioned above. The two peaks at higher binding energies, 533.14\,eV and 531.99\,eV, are attributed to surface contamination such as organic compounds and hydroxides \cite{Biesinger2011, Moulder1992}. These compounds desorb with increasing annealing temperature, as indicated by decreasing intensity of the corresponding peaks.
All peaks (Cu\,2p, W\,4f, O\,1s) in Fig. \ref{fig XPS mix} exhibit subtle but consistent shifts toward lower binding energies with increasing annealing temperature. This behavior can be related to the crystallization of the material at an elevated temperature.


\subsection{Gas sensing measurements}
\label{4pp}

\begin{figure}[h]
    \centerline{\includegraphics[width=1\linewidth]{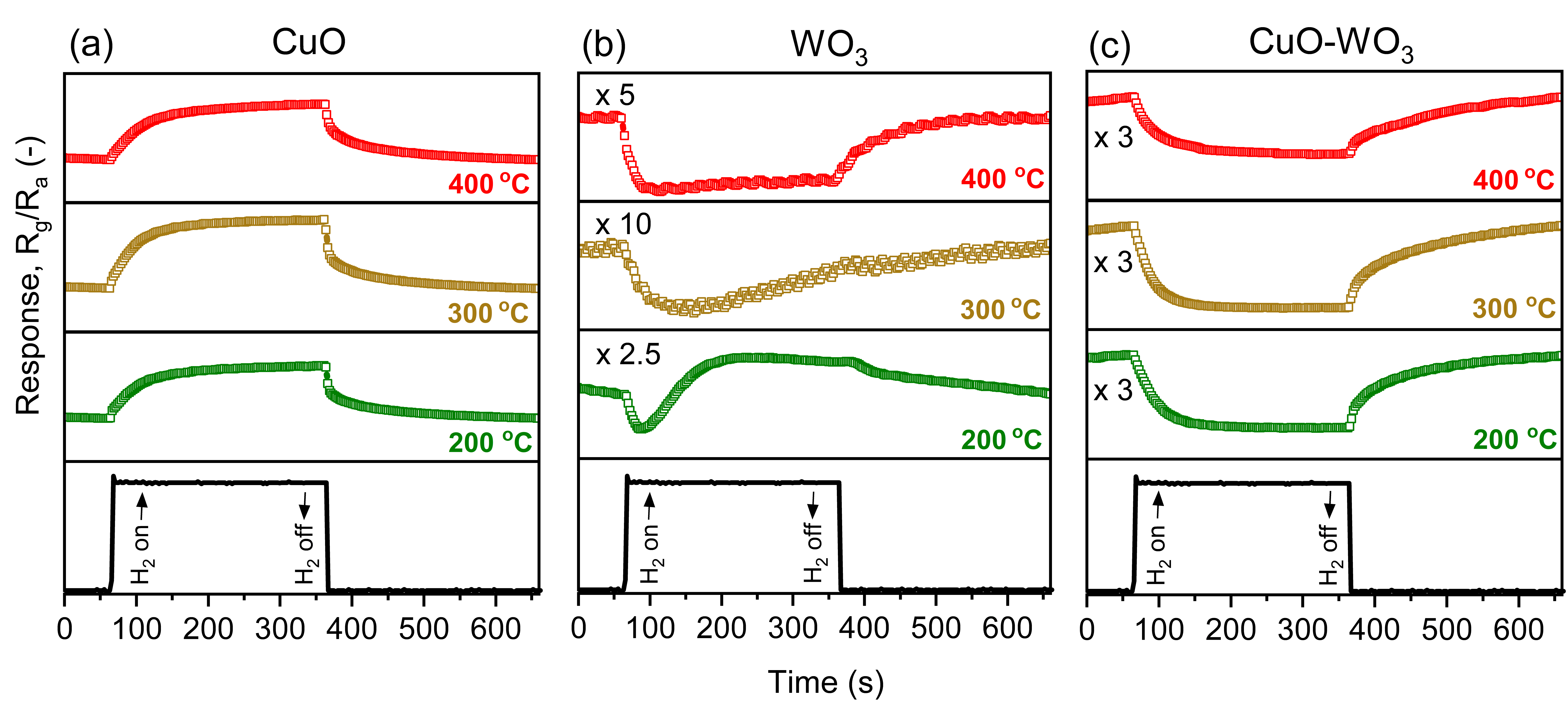}}
            \caption{Dynamic responses to 1\,vol.\% H$_2$ gas at a fixed operating temperature of 200\,$^{\circ}$C for NP-based CuO (a), WO$_3$ (b), and CuO--WO$_3$ (c) films for the as-deposited state and after annealing at different temperatures. The sensing response is defined as R$_g$/R$_a$, where R$_g$ is the resistance in synthetic air containing H$_2$, and R$_a$ is the resistance in synthetic air. The CuO response is shown at its original scale, whereas that of WO$_3$ and CuO–WO$_3$ are magnified by the factors indicated on each curve.}
            \label{fig:4pp}
\end{figure}

The CuO, WO$_3$, and CuO--WO$_3$ films annealed at temperatures of 200, 300, and 400\,$^\circ$C were exposed to 1\% by volume of H$_2$ gas in synthetic air, and their dynamic response characteristics (Fig. \ref{fig:4pp}) were studied at a fixed operating temperature of 200\,$^\circ$C.

All annealed CuO films exhibit p-type behavior, as indicated in Fig.\ref{fig:4pp}a by an increase in resistance upon H$_2$ exposure. This increase is caused by the interaction of H$_2$ with oxygen species  (O$_2^-$, O$^-$, O$^{2-}$) adsorbed from air onto the surface, which causes previously trapped electrons to be released back into the hole accumulation layer of p-type CuO, thereby reducing the hole concentration \cite{Korotcenkov2017}.

The film annealed at 300\,$^\circ$C demonstrates a slightly enhanced response compared to those annealed at 200\,$^\circ$C and 400\,$^\circ$C. This enhanced performance arises from an interplay between the particle size and the crystallinity. Although the film annealed at 200\,$^\circ$C has a larger surface area, its low crystallinity limits the sensitivity, whereas the highly crystalline film annealed at 400\,$^\circ$C exhibits larger NPs and thus a reduced surface area \cite {Katoch2013}. 

The effect of annealing on the sensing response of the WO$_3$ films is particularly interesting and distinctly different from that of conventional n-type WO$_3$, which exhibits a decrease in resistance upon H$_2$ exposure as electrons previously trapped by adsorbed oxygen species are released back into the electron depletion zone of WO$_3$. The most pronounced anomalous behavior can be observed in the WO$_3$ film with the disordered structure annealed at 200\,$^\circ$C. In particular, the resistance initially drops upon exposure to H$_2$, then gradually increases, and eventually stabilizes. When H$_2$ supply is stopped, the resistance slowly decreases, which also mimics the p-type behavior. In the film annealed at 300\,$^\circ$C, where crystallization already occurs, this anomalous response persists but is less pronounced. It becomes even less significant in the film annealed at 400\,$^\circ$C. In these films, the resistance initially drops upon exposure to H$_2$, but then slowly increases instead of the decreasing trend typical for the n-type semiconductors. That is, the crystallinity seems to be a crucial factor controlling the p-type vs. n-type sensing response of the NP-based WO$_3$ films. 

Finally, the CuO--WO$_3$ composite films exhibit an n-type sensing response at all annealing temperatures, as evidenced by their resistance decreasing upon H$_2$ exposure. The response is very similar for the disordered films annealed at 200\,$^{\circ}$C and 300\,$^{\circ}$C, while the crystalline film with the $\gamma$-CuWO$_4$ phase, annealed at 400\,$^{\circ}$C, shows a slightly reduced response to H$_2$. However, none of the composite films demonstrate a significant enhancement in H$_2$ sensing compared to the individual binary films, contrary to our initial expectations.

\section{Discussion}

NP growth, as observed in all investigated CuO, WO$_3$, and CuO–WO$_3$ films, is primarily driven by the tendency of the system to minimize the overall surface energy. Thermal annealing provides sufficient thermal energy to overcome the activation energy for bond breaking and subsequent atomic migration \cite{Bechelany2010, Chen2013, Aljawfi2020}. Two primary mechanisms are commonly associated with the growth of NPs during annealing: coalescence and Ostwald ripening \cite{Grillo2017}.

In dense NP-based materials like those in our study, the predominant growth mode is coalescence. In this process, the growth occurs at relatively low energies via surface migration of atoms and grain boundary diffusion between NPs in contact. With sufficient energy, this may result in the formation of a single, larger NP. In contrast, Ostwald ripening typically involves atoms detaching from smaller NPs and migrating to larger ones, which leads to a redistribution of material without direct particle contact. This process causes smaller NPs to shrink, while larger ones grow as a result of differences in surface energy. However, Ostwald ripening is more common in well-separated systems, where higher temperatures than those used in this study are required to facilitate atom evaporation and reattachment \cite{Li2020, Grillo2017}.

During coalescence, when two NPs come into contact, high-energy, unstable surface atoms (with few neighbors) rapidly diffuse to the more stable interface or point of contact (with more neighbors) \cite{Grammatikopoulos2014}. The formation of new bonds at the interface and the annihilation of free surfaces produce heat sufficient to temporarily 'melt' the interface and form a neck; 'melting' here refers to thermally induced atomic disordering and bond softening that allow atoms to rearrange into a lower-energy configuration \cite{Grammatikopoulos2019}. In amorphous NPs, this triggers a 'crystallization wave' propagating through them as heat dissipates, which results in full crystallization \cite{Jose2005, Grammatikopoulos2014}. Thus, coalescence may promote the crystallization of amorphous particles, and the entire process is accelerated at higher temperatures due to enhanced atomic migration \cite{Aljawfi2020, Jundale2012, She2024}.

When fully crystalline NPs come into contact, their crystallographic orientations are often misaligned due to random positioning. At lower temperatures, NPs reorient to maximize the contact area and thus enhance interfacial stability \cite{Grammatikopoulos2019}. Initial contact leads to free surface annihilation via surface diffusion and the formation of a neck between particles. Disordered interface atoms gain energy from heat released during bond formation, which enables rearrangement to align with the crystallographic orientation of one NP \cite{Jose2005}. In the case of incomplete alignment, grain boundaries form at the interface due to mismatched crystal planes \cite{Jose2005, Aljawfi2020, Zhang2012}. Neck broadening, driven by surface and grain boundary diffusion, stabilizes the interface, often resulting in a dumbbell-shaped structure with a grain boundary at the interface. At elevated temperatures, enhanced atomic mobility driven by surface and grain boundary diffusion can facilitate complete coalescence through extensive atomic rearrangements and epitaxial alignment. However, when the misalignment between interacting NPs is substantial, the process may be hindered by the formation of stable grain boundaries due to kinetic and thermodynamic constraints that limit the extent of atomic reorganization. As a result, metastable configurations often emerge at the grain boundaries, which remain energetically stable over extended periods \cite{Grammatikopoulos2019, Jose2005}.

Our study reveals distinct crystallization behaviors for the CuO, WO$_3$, and CuO--WO$_3$ films upon thermal annealing. The CuO films exhibit crystallization at relatively low annealing temperatures (200\,$^{\circ}$C), with significant NP growth evident in the SEM micrographs, particularly at 300\,$^{\circ}$C. In contrast, the onset of crystallization in the WO$_3$ films is delayed, and the increase in particle size is less pronounced compared to CuO. In the composite films, the XRD patterns show broad peaks from the as-deposited state up to 350\,$^{\circ}$C, with the transition to a novel crystalline $\gamma$-CuWO$_4$ phase at 400\,$^{\circ}$C. This transition is accompanied by minimal changes in particle size.

The ease in crystallization of CuO at relatively low temperatures can be attributed to its comparatively low enthalpy of atomization (7.71\,eV/atom). This low enthalpy facilitates bond breaking and enables sufficient atomic migration, even at lower annealing temperatures. At 200\,$^{\circ}$C, the available thermal energy is sufficient to break atomic bonds, thus promoting surface diffusion and the rearrangement of atoms into lower-energy configurations. As a result, early crystallization occurs, which is followed by the growth of NPs. The improving crystallinity with increasing annealing temperature in the CuO films is reflected in the intensifying diffraction peaks corresponding to m-CuO. This observation is further supported by the vibrational modes in the Raman spectrum, which correspond to m-CuO. Furthermore, the decrease in FWHM of the diffraction peaks with annealing suggests an increase in crystallite size due to an enhanced coalescence. Notably, the negligible difference between the crystallite size calculated from FWHM of the (111) diffraction peak and the particle size estimated from SEM micrographs (at 300, 350, and 400\,$^{\circ}$C) implies that the particles are predominantly single-crystalline \cite{Aljawfi2020}. In addition, XPS analysis reveals that the fraction of Cu\ui{2+} states increases at the expense of Cu\ui{1+} with annealing, which is related to an enhanced oxygen incorporation and the progressive formation of stoichiometric CuO.

Compared to the CuO films, the growth of NPs in the WO$_3$ films is significantly limited to the growth observed after annealing at 400\,$^{\circ}$C, as evidenced by SEM micrographs. This slower growth can be attributed to the much higher enthalpy of atomization of WO$_3$ (25.29\,eV/atom), which is three times higher than that of CuO. Consequently, more energy is required for atoms to break their bonds, diffuse and promote neck formation, atomic rearrangement and crystallization. The XRD patterns of the WO$ _3$ films annealed up to 300\,$^{\circ}$C show broad peaks, which indicate their small crystallite sizes and significant structural disorder. At 300\,$^{\circ}$C, the energy supplied is already sufficient for atoms to overcome bonding constraints resulting in the crystallization of WO$_3$ to monoclinic phase. Further annealing does not significantly increase the intensity of the diffraction peaks, which may indicate a limited improvement in crystallinity. However, Raman spectroscopy of the WO$_3$ films show that the intensity of vibrational modes associated with the monoclinic phase increases at and above 300\,$^{\circ}$C, which suggests an enhanced crystallinity. This apparent discrepancy in the results arises from the higher surface sensitivity of Raman spectroscopy compared to XRD \cite{Akgul2014}. Furthermore, XPS analysis reveals that the WO$_3$ films are nearly fully stoichiometric, with only minor substoichiometry in the as-deposited state and after annealing at 200\,$^{\circ}$C. Increasing the annealing temperature leads to oxidation toward fully stoichiometric WO$_3$ and improved crystallization, as indicated by the shift in the W\ui{6+} peak to lower binding energies above 300\,$^{\circ}$C.

In the CuO--WO$_3$ composite films, a highly disordered atomic arrangement is supposed to occur at the interface between CuO and WO$_3$ NPs, where a grain boundary forms due to the significant difference in the crystallographic orientations of the two types of NPs \cite{Jose2005}. This grain mismatch, which only results in incomplete coalescence, may explain the broad diffraction features observed in the films up to 350\,$^{\circ}$C. Furthermore, the composite film comprises alternating monolayers of CuO and WO$_3$ NPs in equal volumetric proportions, which, however, does not correspond to an inherently equal ratio of Cu and W atoms due to the differences in the atomic volumes of CuO and WO$_3$. CuO with a smaller atomic volume (approximately 21\,\AA$^3$ per metal atom) allows more Cu atoms to be packed into the same unit volume compared to WO$_3$ (approximately 54\,\AA$^3$ per metal atom). This Cu-rich environment may then facilitate the incorporation of Cu atoms into the W sublattice, thereby relieving local lattice strain and promoting the stabilization of the metastable $\gamma$-CuWO$_4$ phase at the grain boundary as the annealing temperature rises to 400\,$^{\circ}$C \cite{Yassine2016, Novikov2015}. This is supported by XRD findings and further corroborated by the Raman and XPS analyses.

The gas-sensing measurements of all three NP-based films clearly reveal that the observed microstructural evolution is a key factor governing their sensing response. The coalescence of NPs, along with their size, crystallinity, and phase composition, directly influences the adsorption and reaction dynamics on the film surfaces upon H$_2$ exposure.

In the case of p-type CuO films, while the enhanced crystallinity with increasing annealing temperature is generally beneficial, it is counteracted by the pronounced growth of NPs with a reduction in surface area. As a result, optimal sensing performance is achieved at 300\,$^{\circ}$C when a delicate balance between the crystallinity and the surface area is maintained.

In contrast, the much less pronounced growth of NPs in the WO$_3$ films upon annealing suggests that the crystallinity enhancement is the dominant factor affecting their sensing behavior. The anomalous sensing response, characterized by a combination of p-type and n-type behavior, observed in these films can be attributed to complex surface interactions between the surface of WO$_3$, oxygen species, and hydrogen gas, which may depend on the structural disorder as proposed in \cite{Maity2015, Pati2014, Yu2023}. The low crystallinity and thus the higher degree of structural disorder of the films annealed at lower temperatures might favor the adsorption of molecular oxygen (O$_2^-$) rather than the dissociation of O--O bonds and the subsequent adsorption of atomic oxygen (O$^-$). When H$_2$ reacts with O$_2^-$, it depletes electrons in WO$_3$ forming hydroxide ions (OH$^-$) and the resistance increases. On the other hand, the reaction of H$_2$ with O$^-$ generates H$_2$O molecules while releasing electrons, thereby reducing resistance. As the crystallinity improves with increasing annealing temperature, the latter reaction becomes dominant, causing the sensing response to gradually shift toward the typical n-type behavior of WO$_3$.

The n-type CuO--WO$_3$ composite films showed the highest thermal stability of the microstructure among all the investigated films, with no noticeable NP growth or crystallization up to 350\,$^{\circ}$C. We attribute this behavior to the effect of the alternating CuO and WO$_3$ NPs monolayers, where the coalescence of NPs and their growth is limited, as the CuO NPs are adjacent to the WO$_3$ NPs from above and bottom and vice versa. Consequently, their sensing response remains largely unchanged and no significant enhancement is observed. At 400\,$^{\circ}$C, the crystallization into the novel $\gamma$-CuWO$_4$ phase occurs, but even this transformation does not result in enhanced sensing performance.  The lack of the sensing enhancement compared to the individual binary films may be due to the nearly identical volumetric ratio of CuO and WO$_3$ NPs within the composite films, which could restrict the synergistic effects typically observed in heterojunction-based sensors \cite{Yang2021, Haviar2018}. Our very recent results indicate that optimizing the CuO/WO$_3$ volumetric ratio could yield better results. A more systematic investigation into this optimization, as well as the sensing response of the $\gamma$-CuWO$_4$ phase, is currently underway and will be presented in a forthcoming publication.

\section{Conclusions}

NP-based CuO, WO$_3$, and their composite CuO-WO$_3$ films were reactively synthesized using a magnetron-based gas aggregation source controlled by an advanced in-house-built software. The films were subsequently subjected to thermal annealing in air up to 400\,$^\circ$C to systematically investigate the microstructure evolution from the atomic to the morphological scale. The thermally-induced changes in the films, including coalescence of NPs, their size growth, crystallinity, and phase composition were analyzed in relation to their gas-sensing response to H$_2$ exposure.

The CuO films exhibit a gradual enhancement of the crystallinity with increasing annealing temperature and a pronounced NP growth occurring at 300\,$^\circ$C and above. The optimal gas-sensing performance of these films, characterized by the p-type behavior, is achieved at 300\,$^\circ$C due to a delicate balance between the particle size (affecting surface area) and the crystallinity. In contrast, WO$_3$ films begin to crystallize only at 300\,$^{\circ}$C and particle growth occurs at 400\,$^{\circ}$C. Their gas-sensing response is anomalous, showing both p-type and n-type characteristics, which gradually shifts toward the typical n-type behavior as the annealing temperature increases. This anomaly is attributed to complex interactions of the films with different oxygen species, where their adsorption is closely linked to the degree of the structural disorder. The composite CuO-WO$_3$ films, consisting of alternating CuO and WO$_3$ monolayers, demonstrate the highest thermal stability of the microstructure. They remain stable across the annealing range and crystallize only at 400\,$^{\circ}$C into a novel phase, $\gamma$-CuWO$_4$, which structurally resembles the $\gamma$-CuMoO$_4$ phase. The gas-sensing response of these films follow the n-type behavior and directly reflects the changes in the microstructure. Interestingly, the disordered structure shows superior gas-sensing performance compared to crystalline $\gamma$-CuWO$_4$. 

The findings provide deeper insights into the thermally induced effects in NP-based films, particularly in relation to their gas-sensing performance. This knowledge will serve as a valuable foundation for future research. However, the anticipated synergistic effects of heterojunctions in the composite films, which were expected to enhance sensitivity to H$_2$, were not observed in this study. Future efforts will thus focus on optimizing the volumetric ratio of CuO and WO$_3$ NPs to fully harness the synergistic potential of nano-heterojunctions and improve the gas sensing performance of the composite CuO-WO$_3$ films.

\section*{Acknowledgement}
This work was supported by the project Quantum materials for applications in sustainable technologies (QM4ST), funded as project No. CZ.02.01.01/00/22\_008/0004572 by Programme Johannes Amos Comenius, call Excellent Research. The authors sincerely thank Prof. Ján Minár and Dr. Laurent Nicola{\" i} from New Technologies–Research Centre, University of West Bohemia in Pilsen, and Dr. Klára Beranová from the Institute of Physics of the Czech Academy of Sciences for their invaluable assistance in interpreting the XPS spectra.

\printcredits

\section*{Declaration of generative AI and AI-assisted technologies in the writing process}

During the preparation of this work the authors used ChatGPT in order to improve the grammar and style of the draft text. After using this service, the authors reviewed and edited the content as needed and take full responsibility for the content of the publication.

\bibliography{biblio}

\begin{thebibliography}{10}
\expandafter\ifx\csname url\endcsname\relax
  \def\url#1{\texttt{#1}}\fi
\expandafter\ifx\csname urlprefix\endcsname\relax\def\urlprefix{URL }\fi
\expandafter\ifx\csname href\endcsname\relax
  \def\href#1#2{#2} \def\path#1{#1}\fi

\bibitem{Manoharan2019}
Y.~Manoharan, S.~E. Hosseini, B.~Butler, H.~Alzhahrani, B.~T.~F. Senior, T.~Ashuri, J.~Krohn, {Hydrogen Fuel Cell Vehicles; Current Status and Future Prospect}, Applied Sciences 9 (2019) 2296.
\newblock \href {https://doi.org/10.3390/app9112296} {\path{doi:10.3390/app9112296}}.

\bibitem{Muradov2008}
N.~MURADOV, T.~VEZIROGLU, “green” path from fossil-based to hydrogen economy: An overview of carbon-neutral technologies, International Journal of Hydrogen Energy 33 (2008) 6804--6839.
\newblock \href {https://doi.org/10.1016/j.ijhydene.2008.08.054} {\path{doi:10.1016/j.ijhydene.2008.08.054}}.

\bibitem{Okazaki2003}
S.~Okazaki, H.~Nakagawa, S.~Asakura, Y.~Tomiuchi, N.~Tsuji, H.~Murayama, M.~Washiya, {Sensing characteristics of an optical fiber sensor for hydrogen leak}, Sensors and Actuators B: Chemical 93 (2003) 142--147.
\newblock \href {https://doi.org/10.1016/S0925-4005(03)00211-9} {\path{doi:10.1016/S0925-4005(03)00211-9}}.

\bibitem{Antonio2022}
A.~Vázquez-López, J.~Bartolomé, A.~Cremades, D.~Maestre, {High-Performance Room-Temperature Conductometric Gas Sensors: Materials and Strategies}, Chemosensors 10 (2022) 227.
\newblock \href {https://doi.org/10.3390/chemosensors10060227} {\path{doi:10.3390/chemosensors10060227}}.

\bibitem{Joshi2018}
N.~Joshi, T.~Hayasaka, Y.~Liu, H.~Liu, O.~N. Oliveira, L.~Lin, {A review on chemiresistive room temperature gas sensors based on metal oxide nanostructures, graphene and 2D transition metal dichalcogenides}, Microchimica Acta 185 (2018) 213.
\newblock \href {https://doi.org/10.1007/s00604-018-2750-5} {\path{doi:10.1007/s00604-018-2750-5}}.

\bibitem{Goel2023}
N.~Goel, K.~Kunal, A.~Kushwaha, M.~Kumar, {Metal oxide semiconductors for gas sensing}, Engineering Reports 5 (6 2023).
\newblock \href {https://doi.org/10.1002/eng2.12604} {\path{doi:10.1002/eng2.12604}}.

\bibitem{Nadargi2023}
D.~Y. Nadargi, A.~Umar, J.~D. Nadargi, S.~A. Lokare, S.~Akbar, I.~S. Mulla, S.~S. Suryavanshi, N.~L. Bhandari, M.~G. Chaskar, {Gas sensors and factors influencing sensing mechanism with a special focus on MOS sensors}, Journal of Materials Science 58 (2023) 559--582.
\newblock \href {https://doi.org/10.1007/s10853-022-08072-0} {\path{doi:10.1007/s10853-022-08072-0}}.

\bibitem{Zhao2023}
H.~Zhao, Y.~Wang, Y.~Zhou, {Accelerating the Gas–Solid Interactions for Conductometric Gas Sensors: Impacting Factors and Improvement Strategies}, Materials 16 (2023) 3249.
\newblock \href {https://doi.org/10.3390/ma16083249} {\path{doi:10.3390/ma16083249}}.

\bibitem{Girolamo2009}
G.~D. Francia, B.~Alfano, V.~L. Ferrara, {Conductometric Gas Nanosensors}, Journal of Sensors 2009 (1 2009).
\newblock \href {https://doi.org/10.1155/2009/659275} {\path{doi:10.1155/2009/659275}}.

\bibitem{Ghenadii}
G.~Korotcenkov, G.~Korotcenkov, {Current Trends in Nanomaterials for Metal Oxide-Based Conductometric Gas Sensors: Advantages and Limitations. Part 1: 1D and 2D Nanostructures.}, nanomaterials 10 (2020) 1392.
\newblock \href {https://doi.org/10.3390/nano10071392} {\path{doi:10.3390/nano10071392}}.

\bibitem{Jun2009}
J.~H. Jun, J.~Yun, K.~Cho, I.-S. Hwang, J.-H. Lee, S.~Kim, {Necked ZnO nanoparticle-based NO$_2$ sensors with high and fast response}, Sensors and Actuators B: Chemical 140 (2009) 412--417.
\newblock \href {https://doi.org/10.1016/j.snb.2009.05.019} {\path{doi:10.1016/j.snb.2009.05.019}}.

\bibitem{Korotcenkov2017}
G.~Korotcenkov, B.~K. Cho, {Metal oxide composites in conductometric gas sensors: Achievements and challenges} (2017).
\newblock \href {https://doi.org/10.1016/j.snb.2016.12.117} {\path{doi:10.1016/j.snb.2016.12.117}}.

\bibitem{Yang2021}
S.~Yang, G.~Lei, H.~Xu, Z.~Lan, Z.~Wang, H.~Gu, {Metal oxide based heterojunctions for gas sensors: A review} (4 2021).
\newblock \href {https://doi.org/10.3390/nano11041026} {\path{doi:10.3390/nano11041026}}.

\bibitem{Steinhauer2021}
S.~Steinhauer, S.~Steinhauer, S.~Steinhauer, {Gas Sensors Based on Copper Oxide Nanomaterials: A Review}, Chemosensors 9 (2021) 51.
\newblock \href {https://doi.org/10.3390/chemosensors9030051} {\path{doi:10.3390/chemosensors9030051}}.

\bibitem{Dong2020}
C.~Dong, R.~Zhao, L.~Yao, Y.~Ran, X.~Zhang, Y.~Wang, {A review on WO$_3$ based gas sensors: Morphology control and enhanced sensing properties} (4 2020).
\newblock \href {https://doi.org/10.1016/j.jallcom.2019.153194} {\path{doi:10.1016/j.jallcom.2019.153194}}.

\bibitem{Kumar2020}
N.~Kumar, J.~Čapek, S.~Haviar, {Nanostructured CuWO$_4$/WO$_{3-x}$ films prepared by reactive magnetron sputtering for hydrogen sensing}, International Journal of Hydrogen Energy 45 (2020) 18066--18074.
\newblock \href {https://doi.org/10.1016/j.ijhydene.2020.04.203} {\path{doi:10.1016/j.ijhydene.2020.04.203}}.

\bibitem{Kumar2021}
N.~Kumar, S.~Haviar, P.~Zeman, {Three-layer PdO/CuWO$_4$/CuO system for hydrogen gas sensing with reduced humidity interference}, Nanomaterials 11 (12 2021).
\newblock \href {https://doi.org/10.3390/nano11123456} {\path{doi:10.3390/nano11123456}}.

\bibitem{Haviar2018}
S.~Haviar, J.~Čapek, Šárka Batková, N.~Kumar, F.~Dvořák, T.~Duchoň, M.~Fialová, P.~Zeman, {Hydrogen gas sensing properties of WO$_3$ sputter-deposited thin films enhanced by on-top deposited CuO nanoclusters}, International Journal of Hydrogen Energy 43 (2018) 22756--22764.
\newblock \href {https://doi.org/10.1016/j.ijhydene.2018.10.127} {\path{doi:10.1016/j.ijhydene.2018.10.127}}.

\bibitem{Shaji2024}
K.~Shaji, S.~Haviar, P.~Zeman, Šimon Kos, R.~Čerstvý, J.~Čapek, {Controlled sputter deposition of oxide nanoparticles-based composite thin films}, Surface and Coatings Technology 477 (2024) 130325.
\newblock \href {https://doi.org/10.1016/J.SURFCOAT.2023.130325} {\path{doi:10.1016/J.SURFCOAT.2023.130325}}.

\bibitem{Biesinger2022}
M.~C. Biesinger, {Accessing the robustness of adventitious carbon for charge referencing (correction) purposes in XPS analysis: Insights from a multi-user facility data review}, Applied Surface Science 597 (2022) 153681.
\newblock \href {https://doi.org/10.1016/J.APSUSC.2022.153681} {\path{doi:10.1016/J.APSUSC.2022.153681}}.

\bibitem{Haviar2025}
S.~Haviar, B.~Prifling, T.~Kozák, K.~Shaji, T.~Košutová, Šimon Kos, V.~Schmidt, J.~Čapek, \href{https://linkinghub.elsevier.com/retrieve/pii/S266652392400117X}{{Analysis and 3D modelling of percolated conductive networks in nanoparticle-based thin films}}, Applied Surface Science Advances 25 (2025) 100689.
\newblock \href {https://doi.org/10.1016/j.apsadv.2024.100689} {\path{doi:10.1016/j.apsadv.2024.100689}}.
\newline\urlprefix\url{https://linkinghub.elsevier.com/retrieve/pii/S266652392400117X}

\bibitem{Chai2015}
Y.~Chai, C.~W. Tam, K.~P. Beh, F.~K. Yam, Z.~Hassan, {Effects of thermal treatment on the anodic growth of tungsten oxide films}, Thin Solid Films 588 (2015) 44--49.
\newblock \href {https://doi.org/10.1016/j.tsf.2015.04.033} {\path{doi:10.1016/j.tsf.2015.04.033}}.

\bibitem{CHEN2021}
H.~Chen, A.~Chiasera, S.~Varas, O.~Sayginer, C.~Armellini, G.~Speranza, R.~Suriano, M.~Ferrari, S.~M. Pietralunga, {Tungsten oxide films by radio-frequency magnetron sputtering for near-infrared photonics}, Optical Materials: X 12 (2021) 100093.
\newblock \href {https://doi.org/10.1016/j.omx.2021.100093} {\path{doi:10.1016/j.omx.2021.100093}}.

\bibitem{houska2025newpolymorphgammaCuWO4inspired}
J.~Houska, S.~Haviar, J.~Capek, R.~Cerstvy, K.~Shaji, N.~Kumar, P.~Zeman, \href{https://arxiv.org/abs/2501.03036}{{New polymorph $\gamma$-CuWO$_4$ inspired by $\gamma$-CuMoO$_4$: experimental identification and theoretical verification}} (2025).
\newblock \href {http://arxiv.org/abs/2501.03036} {\path{arXiv:2501.03036}}.
\newline\urlprefix\url{https://arxiv.org/abs/2501.03036}

\bibitem{Akgul2014}
F.~A. Akgul, G.~Akgul, N.~Yildirim, H.~E. Unalan, R.~Turan, {Influence of thermal annealing on microstructural, morphological, optical properties and surface electronic structure of copper oxide thin films}, Materials Chemistry and Physics 147 (2014) 987--995.
\newblock \href {https://doi.org/10.1016/j.matchemphys.2014.06.047} {\path{doi:10.1016/j.matchemphys.2014.06.047}}.

\bibitem{Xu1999}
J.~F. Xu, W.~Ji, Z.~X. Shen, W.~S. Li, S.~H. Tang, X.~R. Ye, D.~Z. Jia, X.~Q. Xin, \href{https://analyticalsciencejournals.onlinelibrary.wiley.com/doi/10.1002/}{{Raman Spectra of CuO Nanocrystals}}, JOURNAL OF RAMAN SPECTROSCOPY 30 (1999) 413--415.
\newblock \href {https://doi.org/10.1002/(SICI)1097-4555(199905)30:5<413::AID-JRS387>3.0.CO;2-N} {\path{doi:10.1002/(SICI)1097-4555(199905)30:5<413::AID-JRS387>3.0.CO;2-N}}.
\newline\urlprefix\url{https://analyticalsciencejournals.onlinelibrary.wiley.com/doi/10.1002/}

\bibitem{Nanba1991}
T.~Nanba, Y.~Nishiyama, I.~Yasui, {Structural study of amorphous WO$_3$ thin films prepared by the ion exchange method}, Journal of Materials Research 6 (1991) 1324--1333.
\newblock \href {https://doi.org/10.1557/JMR.1991.1324} {\path{doi:10.1557/JMR.1991.1324}}.

\bibitem{Santato2001}
C.~Santato, M.~Odziemkowski, M.~Ulmann, J.~Augustynski, {Crystallographically Oriented Mesoporous WO$_3$ Films: Synthesis, Characterization, and Applications}, Journal of the American Chemical Society 123 (2001) 10639--10649.
\newblock \href {https://doi.org/10.1021/ja011315x} {\path{doi:10.1021/ja011315x}}.

\bibitem{Song2019}
W.~Song, W.~Song, R.~Zhang, R.~Zhang, R.~Zhang, R.~Zhang, X.~Bai, X.~Bai, Q.~Jia, Q.~Jia, H.~Ji, H.~Ji, {Exposed crystal facets of WO 3 nanosheets by phase control on NO 2 -sensing performance}, Journal of Materials Science: Materials in Electronics 31 (2019) 610--620.
\newblock \href {https://doi.org/10.1007/s10854-019-02565-6} {\path{doi:10.1007/s10854-019-02565-6}}.

\bibitem{Ruiz‐Fuertes2008}
J.~Ruiz‐Fuertes, J.~Ruiz-Fuertes, D.~Errandonea, D.~Errandonea, A.~Segura, A.~Segura, F.~J. Manjón, F.~J. Manjón, Z.~Zhu, Z.~Zhu, C.~Tu, C.~Y. Tu, {Growth, characterization, and high-pressure optical studies of CuWO$_4$}, High Pressure Research: An International Journal 28 (2008) 565--570.
\newblock \href {https://doi.org/10.1080/08957950802446643} {\path{doi:10.1080/08957950802446643}}.

\bibitem{Ali2022}
N.~U. H.~L. Ali, S.~Manoharan, P.~Pazhamalai, S.-J. Kim, {CuMoO$_4$ nanostructures: A novel bifunctional material for supercapacitor and sensor applications}, Journal of Energy Storage 52 (2022) 104784.
\newblock \href {https://doi.org/10.1016/j.est.2022.104784} {\path{doi:10.1016/j.est.2022.104784}}.

\bibitem{Biesinger2011}
M.~C. Biesinger, B.~P. Payne, A.~P. Grosvenor, L.~W. Lau, A.~R. Gerson, R.~S.~C. Smart, {Resolving surface chemical states in XPS analysis of first row transition metals, oxides and hydroxides: Cr, Mn, Fe, Co and Ni}, Applied Surface Science 257 (2011) 2717--2730.
\newblock \href {https://doi.org/10.1016/j.apsusc.2010.10.051} {\path{doi:10.1016/j.apsusc.2010.10.051}}.

\bibitem{Moulder1992}
J.~F. Moulder, J.~Chastain, {Handbook of x-ray photoelectron spectroscopy : a reference book of standard spectra for identification and interpretation of XPS data}, Physical Electronics Division, Perkin-Elmer Corp., 1992.

\bibitem{Shpak2007}
A.~P. Shpak, A.~M. Korduban, M.~M. Medvedskij, V.~O. Kandyba, {XPS studies of active elements surface of gas sensors based on WO$_{3-x}$ nanoparticles}, Journal of Electron Spectroscopy and Related Phenomena 156-158 (2007) 172--175.
\newblock \href {https://doi.org/10.1016/j.elspec.2006.12.059} {\path{doi:10.1016/j.elspec.2006.12.059}}.

\bibitem{Khyzhun2002}
O.~Y. Khyzhun, Y.~M. Solonin, V.~D. Dobrovolsky, {ELECTRONIC STRUCTURE OF HyWO$_3$ AND WO$_x$ STUDIED BY THE XPS, XES, AND XAS METHODS} (2002).

\bibitem{Liu2018}
H.~Y. Liu, Y.~L. Hsu, H.~Y. Su, R.~C. Huang, F.~Y. Hou, G.~C. Tu, W.~H. Liu, {A comparative study of amorphous, anatase, rutile, and mixed phase TiO$_2$ films by mist chemical vapor deposition and ultraviolet photodetectors applications}, IEEE Sensors Journal 18 (2018) 4022--4029.
\newblock \href {https://doi.org/10.1109/JSEN.2018.2819700} {\path{doi:10.1109/JSEN.2018.2819700}}.

\bibitem{Gajraj2021}
V.~Gajraj, C.~R. Mariappan, {CuWO$_4$: A promising multifunctional electrode material for energy storage as in redox active solid-state asymmetric supercapacitor and an electrocatalyst for energy conversion in methanol electro-oxidation}, Journal of Electroanalytical Chemistry 895 (8 2021).
\newblock \href {https://doi.org/10.1016/j.jelechem.2021.115504} {\path{doi:10.1016/j.jelechem.2021.115504}}.

\bibitem{Khyzhun2005}
O.~Y. Khyzhun, T.~Strunskus, S.~Cramm, Y.~M. Solonin, {Electronic structure of CuWO$_4$: XPS, XES and NEXAFS studies}, Journal of Alloys and Compounds 389 (2005) 14--20.
\newblock \href {https://doi.org/10.1016/j.jallcom.2004.08.013} {\path{doi:10.1016/j.jallcom.2004.08.013}}.

\bibitem{Katoch2013}
A.~Katoch, G.~J. Sun, S.~W. Choi, J.~H. Byun, S.~S. Kim, {Competitive influence of grain size and crystallinity on gas sensing performances of ZnO nanofibers}, Sensors and Actuators, B: Chemical 185 (2013) 411--416.
\newblock \href {https://doi.org/10.1016/j.snb.2013.05.030} {\path{doi:10.1016/j.snb.2013.05.030}}.

\bibitem{Bechelany2010}
M.~Bechelany, X.~Maeder, J.~Riesterer, J.~Hankache, D.~Lerose, S.~Christiansen, J.~Michler, L.~Philippe, {Synthesis Mechanisms of Organized Gold Nanoparticles: Influence of Annealing Temperature and Atmosphere}, Crystal Growth \& Design 10 (2010) 587--596.
\newblock \href {https://doi.org/10.1021/cg900981q} {\path{doi:10.1021/cg900981q}}.

\bibitem{Chen2013}
H.~Chen, Y.~Yu, H.~L. Xin, K.~A. Newton, M.~E. Holtz, D.~Wang, D.~A. Muller, H.~D. Abruña, F.~J. DiSalvo, {Coalescence in the Thermal Annealing of Nanoparticles: An in Situ STEM Study of the Growth Mechanisms of Ordered Pt–Fe Nanoparticles in a KCl Matrix}, Chemistry of Materials 25 (2013) 1436--1442.
\newblock \href {https://doi.org/10.1021/cm303489z} {\path{doi:10.1021/cm303489z}}.

\bibitem{Aljawfi2020}
R.~N. Aljawfi, M.~J. Alam, F.~Rahman, S.~Ahmad, A.~Shahee, S.~Kumar, {Impact of annealing on the structural and optical properties of ZnO nanoparticles and tracing the formation of clusters via DFT calculation}, Arabian Journal of Chemistry 13 (2020) 2207--2218.
\newblock \href {https://doi.org/10.1016/j.arabjc.2018.04.006} {\path{doi:10.1016/j.arabjc.2018.04.006}}.

\bibitem{Grillo2017}
F.~Grillo, H.~V. Bui, J.~A. Moulijn, M.~T. Kreutzer, J.~R. van Ommen, {Understanding and Controlling the Aggregative Growth of Platinum Nanoparticles in Atomic Layer Deposition: An Avenue to Size Selection}, The Journal of Physical Chemistry Letters 8 (2017) 975--983.
\newblock \href {https://doi.org/10.1021/acs.jpclett.6b02978} {\path{doi:10.1021/acs.jpclett.6b02978}}.

\bibitem{Li2020}
M.~Li, A.~Borsay, M.~Dakhchoune, K.~Zhao, W.~Luo, A.~Züttel, {Thermal stability of size-selected copper nanoparticles: Effect of size, support and CO$_2$ hydrogenation atmosphere}, Applied Surface Science 510 (2020) 145439.
\newblock \href {https://doi.org/10.1016/j.apsusc.2020.145439} {\path{doi:10.1016/j.apsusc.2020.145439}}.

\bibitem{Grammatikopoulos2014}
P.~Grammatikopoulos, C.~Cassidy, V.~Singh, M.~Sowwan, {Coalescence-induced crystallisation wave in Pd nanoparticles}, Scientific Reports 4 (4 2014).
\newblock \href {https://doi.org/10.1038/srep05779} {\path{doi:10.1038/srep05779}}.

\bibitem{Grammatikopoulos2019}
P.~Grammatikopoulos, M.~Sowwan, J.~Kioseoglou, {Computational Modeling of Nanoparticle Coalescence} (6 2019).
\newblock \href {https://doi.org/10.1002/adts.201900013} {\path{doi:10.1002/adts.201900013}}.

\bibitem{Jose2005}
M.~José-Yacamán, C.~Gutierrez-Wing, M.~Miki, D.~Q. Yang, K.~N. Piyakis, E.~Sacher, {Surface diffusion and coalescence of mobile metal nanoparticles}, Journal of Physical Chemistry B 109 (2005) 9703--9711.
\newblock \href {https://doi.org/10.1021/jp0509459} {\path{doi:10.1021/jp0509459}}.

\bibitem{Jundale2012}
D.~M. Jundale, P.~B. Joshi, S.~Sen, V.~B. Patil, {Nanocrystalline CuO thin films: Synthesis, microstructural and optoelectronic properties}, Journal of Materials Science: Materials in Electronics 23 (2012) 1492--1499.
\newblock \href {https://doi.org/10.1007/s10854-011-0616-2} {\path{doi:10.1007/s10854-011-0616-2}}.

\bibitem{She2024}
C.~She, J.~Gao, Z.~Wang, S.~Jin, M.~Chen, L.~Song, K.~Chen, {Coalescence of Al$_2$O$_3$/Al, MgO/Mg, and MgO/Al two nanoparticles during combustion}, Applied Surface Science 649 (3 2024).
\newblock \href {https://doi.org/10.1016/j.apsusc.2023.159157} {\path{doi:10.1016/j.apsusc.2023.159157}}.

\bibitem{Zhang2012}
Y.~Zhang, S.~Li, W.~Yan, S.~D. Tse, {Effect of size-dependent grain structures on the dynamics of nanoparticle coalescence}, Journal of Applied Physics 111 (6 2012).
\newblock \href {https://doi.org/10.1063/1.4730773} {\path{doi:10.1063/1.4730773}}.

\bibitem{Yassine2016}
Y.~E. Mendili, J.~F. Bardeau, N.~Randrianantoandro, J.~M. Greneche, F.~Grasset, {Structural behavior of laser-irradiated $\gamma$-Fe$_2$O$_3$ nanocrystals dispersed in porous silica matrix : $\gamma$-Fe$_2$O$_3$ to $\alpha$-Fe$_2$O$_3$ phase transition and formation of $\epsilon$-Fe$_2$O$_3$}, Science and Technology of Advanced Materials 17 (2016) 597--609.
\newblock \href {https://doi.org/10.1080/14686996.2016.1222494} {\path{doi:10.1080/14686996.2016.1222494}}.

\bibitem{Novikov2015}
V.~Y. Novikov, \href{http://dx.doi.org/10.1016/j.matlet.2015.07.092}{{Grain growth in nanocrystalline materials}}, Materials Letters 159 (2015) 510--513.
\newblock \href {https://doi.org/10.1016/j.matlet.2015.07.092} {\path{doi:10.1016/j.matlet.2015.07.092}}.
\newline\urlprefix\url{http://dx.doi.org/10.1016/j.matlet.2015.07.092}

\bibitem{Maity2015}
A.~Maity, A.~Ghosh, S.~B. Majumder, {Understanding the anomalous conduction behavior in 'n' type tungsten oxide thin film during hydrogen gas sensing: Kinetic analyses of conductance transients}, Sensors and Actuators, B: Chemical 220 (2015) 949--957.
\newblock \href {https://doi.org/10.1016/j.snb.2015.06.038} {\path{doi:10.1016/j.snb.2015.06.038}}.

\bibitem{Pati2014}
S.~Pati, P.~Banerji, S.~Majumder, {n- to p- type carrier reversal in nanocrystalline indium doped ZnO thin film gas sensors}, International Journal of Hydrogen Energy 39 (2014) 15134--15141.
\newblock \href {https://doi.org/10.1016/j.ijhydene.2014.07.075} {\path{doi:10.1016/j.ijhydene.2014.07.075}}.

\bibitem{Yu2023}
Z.~Yu, S.~Lv, Q.~Yao, N.~Fang, Y.~Xu, Q.~Shao, C.~W. Pao, J.~F. Lee, G.~Li, L.~M. Yang, X.~Huang, {Low-Coordinated Pd Site within Amorphous Palladium Selenide for Active, Selective, and Stable H$_2$O$_2$ Electrosynthesis}, Advanced Materials 35 (2 2023).
\newblock \href {https://doi.org/10.1002/adma.202208101} {\path{doi:10.1002/adma.202208101}}.

\end{thebibliography}
\bibliographystyle{elsarticle-num}
%\vskip3pt

\end{document}